\documentclass[alpha-refs]{wiley-article}

\usepackage{siunitx}          % SI units
\usepackage{booktabs}         % \toprule / \midrule / \bottomrule
\usepackage{threeparttable}   % table notes
\usepackage{tabularx}
\usepackage{array}
\usepackage{multirow}
\usepackage{rotating}         % sideways tables (Tab 1)
\usepackage{amsmath}
\usepackage{microtype}
\usepackage{xcolor}
\usepackage{caption}
\papertype{Research Article}
\paperfield{Physiological Ecology}

\title{Body size predicts how long ant workers live --- but not how they age
  or how they die from heat}

\abbrevs{%
  KM, Kaplan--Meier;
  PH, proportional hazards;
  HR, hazard ratio;
  CI, confidence interval;
  AIC, Akaike information criterion;
  LRT, likelihood-ratio test;
  IQR, interquartile range;
  ROS, reactive oxygen species%
}

\author[1]{Alana Moscardi}
\author[2]{Rafael da Silva}
\author[3]{Gleycon Silva}

\affil[1]{Graduate Programme in Ecology, Instituto Nacional de Pesquisas
  da Amaz\^{o}nia (INPA), Manaus, Brazil}
\affil[2]{Graduate Programme in Data Science, Eastern University,
  St.\ Davids, PA, USA}
\affil[3]{Graduate Programme in Ecology, Instituto Nacional de Pesquisas
  da Amaz\^{o}nia (INPA), Manaus, Brazil}

\corraddress{Alana Moscardi, Graduate Programme in Ecology, Instituto
  Nacional de Pesquisas da Amaz\^{o}nia (INPA), Manaus, AM, Brazil}
\corremail{alanamoscardi1@gmail.com}

\runningauthor{Moscardi et al.}
\fundinginfo{[To be completed before submission]}

\begin{document}
\begin{frontmatter}
\maketitle

%% ------ Abstract -------------------------------------------
\begin{abstract}
\small
\begin{enumerate}
  \item In social insects, `mortality risk' bundles three components that
    need not share a common predictor: \emph{duration} (how long a worker lives
    on average), \emph{senescence trajectory} (whether mortality accelerates
    with age, captured by the Weibull shape parameter $\rho$), and
    \emph{thermal vulnerability} (how strongly mortality hazard rises with
    temperature).

  \item We test all three axes in 18 Australian ant species via paired
    field--laboratory survival assays (2,363 cohort-day observations;
    1,148 workers); no single predictor is detectable across all three axes at
    this sample size.

  \item Body size predicts \emph{duration} (Cox HR\,=\,0.67, $p = 0.002$);
    colony size ($p = 0.60$) and the size\,$\times$\,temperature interaction
    ($p = 0.72$) show no detectable moderating effect.  The
    body-size\,$\times$\,foraging-rate interaction is weak but significant
    (LRT $p = 0.014$).  By exclusion of the two null extrinsic moderators,
    intrinsic physiology remains the most parsimonious explanation for the
    main size--longevity pattern, though the foraging-rate interaction
    warrants cautious interpretation (VIF\,$\approx$\,4) and unmeasured
    confounders cannot be ruled out at $N = 18$ species.

  \item \emph{Senescence trajectory} is associated with circadian niche, not size:
    steepest in matinal species (Kruskal--Wallis $p = 0.009$; matinal
    vs.\ crepuscular $p = 0.002$), and uncorrelated with body mass
    (Spearman $p = 0.32$).

  \item \emph{Thermal hazard} plateaus above $T^* = 20$\,\textdegree C
    ($\Delta$AIC\,=\,$-38$; $p < 0.001$); elevated thermal sensitivity
    observed in the \textit{Rhytidoponera} lineage (Ectatomminae) above the
    plateau ($5\,\%$/\textdegree C, $p = 0.015$).

  \item Circadian regime and lineage identity --- not body size --- emerge
    as the climate-relevant axes, although they are strongly collinear in this
    assemblage (Cram\'{e}r's $V = 0.85$) and cannot be fully disentangled
    here; they converge on matinal \textit{Rhytidoponera} (Ectatomminae)
    as the candidate most-vulnerable lineage.  Because hazard plateaus near the
    assemblage mean, the operative climate variable is the frequency of
    extreme-heat days rather than mean warming.  The practical implication is
    concrete and testable: size-based vulnerability indices, which capture only
    the \emph{duration} axis, will systematically misrank the most exposed taxa
    in assemblages where circadian niche and thermal sensitivity are decoupled
    from body size --- a pattern documented here and expected wherever matinal
    lineages co-occur with larger, thermally buffered species.
\end{enumerate}


\end{abstract}
\end{frontmatter}

%% ============================================================
%%  GRAPHICAL ABSTRACT  (optional — NOT a Functional Ecology requirement)
%%  ----------------------------------------------------------------------
%%  Functional Ecology has no graphical-abstract slot. It uses the numbered
%%  Summary above, plus a 250-350 word lay/Plain-Language Summary requested
%%  at the revision stage, and invites a cover photo on acceptance. Keep the
%%  graphical-abstract asset for promotion / a cover submission / the lay
%%  summary. If you later retarget a journal that DOES use graphical
%%  abstracts, uncomment the block below and supply graphical_abstract.pdf.
%% ----------------------------------------------------------------------
% \begin{figure}[htbp]
%   \centering
%   \includegraphics[width=\textwidth]{graphical_abstract}
%   \caption*{Graphical abstract.}
% \end{figure}
%% ============================================================

%% ============================================================
\section{Introduction}
\label{sec:intro}

Body size is among the most intensively studied life-history traits in
insects, and across social insects larger workers generally live longer ---
a pattern documented directly in ants \citep{Riskas2026}.
This robust size--longevity relationship is so well established that body size
is often used as a one-dimensional proxy for mortality risk in comparative and
applied work.
But `mortality risk' bundles together properties that need not share the
same predictor: how \emph{long} a worker lives on average, how
\emph{steeply} its risk of death accelerates with age, and how
\emph{vulnerable} it is to environmental extremes such as heat.  If these
components have distinct drivers, then a size-centric framework will systematically
mispredict which workers --- and which lineages --- are most at risk under
environmental change.

\citet{Riskas2026} documented a strong size--longevity effect in this same
assemblage of 18 Australian ant species, but did not partition mortality risk
into its component axes, test whether size buffers thermal mortality, or
examine the trajectory of senescence.  The present study extends that work by
addressing three questions that are jointly sufficient to test whether body
size is a strong comparative predictor of ant mortality risk in the broad
sense, or only its most commonly measured component.  We organise the study
around three questions, each resolved by a falsifiable claim
(Table~\ref{tab:roadmap}).

\textbf{Q1 --- What kind of mechanism links body size to longevity?}  Three
non-exclusive hypotheses dominate the literature: (i) an
\emph{intrinsic-physiology} mechanism, in which larger workers have lower
mass-specific metabolic rates \citep{West2001, Brown2004} and plausibly
reduced oxidative stress \citep{Kramer2021}; (ii) an
\emph{extrinsic-buffering} mechanism, in which greater mass buffers workers
against desiccation \citep{Lighton1994, Bujan2016} and heat stress
\citep{Baudier2018}; and (iii) a
\emph{colony-level dilution} hypothesis: that larger colonies weaken the
per-worker size benefit.  Discriminating among these
mechanisms requires testing whether the size benefit is modulated by colony
size, foraging context, or temperature --- questions unaddressed by prior
work on this assemblage.

\textbf{Q2 --- What shapes the \emph{trajectory} of mortality?}  Unlike mean
lifespan, the Weibull shape parameter $\rho$ captures \emph{whether}
mortality accelerates with age (senescence, $\rho > 1$) or decelerates
(frailty selection, $\rho < 1$) --- a property that is, at least in
principle, independent of average longevity
\citep{Ricklefs2010, Dahlgren2016, Lemaitre2024}.  Because trajectory
integrates an organism's physiological and ecological context, we ask whether
it tracks body size --- as a naïve extension of the size--longevity rule
would predict --- or instead is associated with circadian activity regime, i.e.\ the
timing of foraging relative to the daily temperature cycle.

\textbf{Q3 --- How does temperature act, and is thermal risk taxon-specific?}
Climate change is increasing global mean temperatures and the frequency of
extreme heat events; in south-eastern Australia specifically, these trends
are well-documented \citep{Herold2021, BoM2025, IPCC2021, PerkinsKirkpatrick2020}.  Whether
thermal mortality scales linearly with temperature, and whether all ant
lineages are equally vulnerable or vulnerability is structured by lineage
identity, is central to predicting community-level responses to warming yet
is rarely quantified at the subfamily level; existing thermal surveys
operate at the species level \citep{Diamond2017, Kaspari2015}.

\begin{figure}[htbp]
\centering
\includegraphics[width=0.65\textwidth]{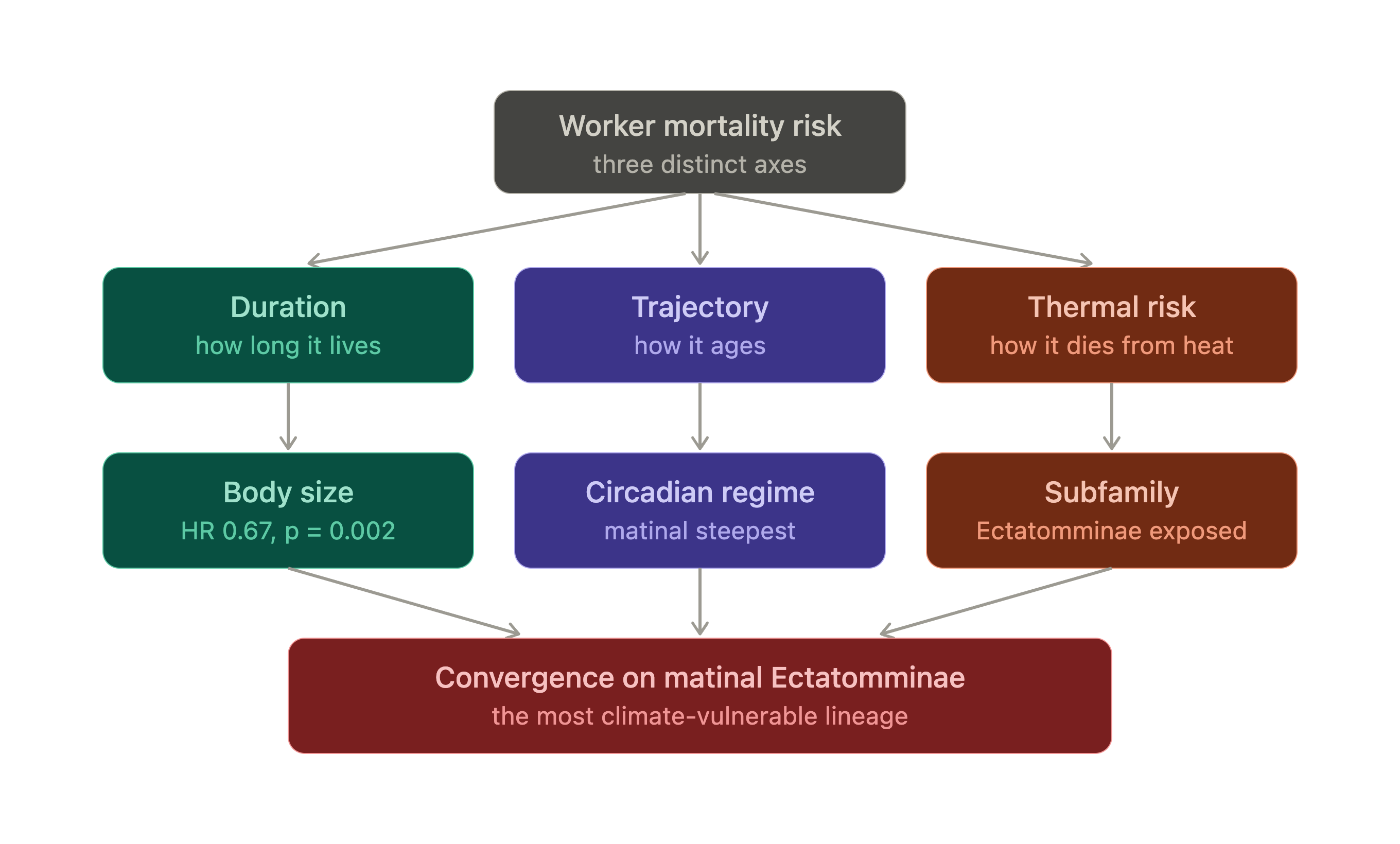}
\caption{\textbf{Conceptual framework: three decoupled axes of worker
  mortality risk.}  Body size predicts only the \emph{duration} of life
  (HR\,=\,0.67); it is statistically decoupled from the \emph{senescence
  trajectory} ($\rho$--mass Spearman $p = 0.32$) and from \emph{thermal
  vulnerability} (size\,$\times$\,temperature $p = 0.72$).  The trajectory
  axis is associated with circadian activity regime (matinal species senesce fastest;
  Kruskal--Wallis $p = 0.009$), and the thermal axis is lineage-patterned
  (elevated thermal sensitivity in the \textit{Rhytidoponera} lineage
  (Ectatomminae); PGLS $p = 0.081$, corroborated by
  individual-level Cox $p = 0.015$).
  Circadian regime and lineage identity are strongly collinear in this assemblage
  (Cram\'{e}r's $V = 0.85$) and cannot be fully separated at $N = 18$
  species; they are therefore drawn as complementary, not independent,
  predictors.  The three climate-relevant signals converge on matinal
  \textit{Rhytidoponera} (Ectatomminae).  Each axis is labelled in text
  on the schematic so the diagram is interpretable independently of colour.
  Abbreviations: HR, hazard ratio (Cox proportional-hazards model);
  $\rho$, Weibull shape parameter (senescence rate); KW, Kruskal--Wallis test;
  PGLS, phylogenetic generalised least squares; Cram\'{e}r's $V$, association
  statistic between two categorical variables.
  \label{fig:synthesis}}
\end{figure}

Body size is a strong comparative predictor of worker longevity, but its
explanatory scope across other axes of mortality risk remains uncertain.
By answering these three questions within a single, unified framework on one
paired field--laboratory system, we are able to test not only each component
in isolation but also whether they share a common predictor --- and we find no
shared predictor detectable at $N = 18$ species.  Figure~\ref{fig:synthesis}
summarises the conceptual framework.

%% ---- Conceptual synthesis schematic (three decoupled axes) -------

%% ============================================================
\section{Material and Methods}
\label{sec:methods}

Each of the three claims is evaluated through a set of falsifiable sub-tests
(H1--H9), described in full below and summarised, with outcomes, in
Table~\ref{tab:roadmap}.  In brief, \textbf{Claim~1} (\S\ref{sec:cox}) asks
whether the size--longevity benefit is genuine and, if so, whether it is
modulated by colony size, foraging rate, or temperature --- the pattern that
distinguishes an intrinsic-physiology mechanism from extrinsic buffering or
colony-level dilution.  \textbf{Claim~2} (\S\ref{sec:weibull}) asks whether
the \emph{shape} of the mortality trajectory (the Weibull senescence rate
$\rho$) is associated with circadian activity regime rather than body size.
\textbf{Claim~3} (\S\ref{sec:thermal}) characterises the
temperature--hazard function --- its shape, and whether thermal
vulnerability is concentrated in particular lineages.  Throughout, the
internal labels H1--H9 cross-reference the methods, tables, supplementary
material, and analysis code.

\subsection{Levels of inference}
\label{sec:levels}

Analyses operate at three nested levels, and results from different levels
are not treated as commensurate without qualification.  \textbf{Cohort-day
(individual survival process)} ($N = 2{,}363$ observations; 1{,}826 events;
106 cohorts): Cox models M0/M1/M2, the piecewise temperature spline, and
subfamily\,$\times$\,temperature interactions.  \textbf{Colony}
($N = 39$): OLS tests of the body-size\,$\times$\,colony-size moderator
(H2a).  \textbf{Species} ($N = 18$, or $N = 14$--15 after phylogenetic or
cohort filters): Weibull $\rho$, Kruskal--Wallis tests of circadian niche,
Spearman correlations, PGLS, and variance partitioning of $\rho$.  Each
claim states its primary inferential unit; species-level tests are
underpowered relative to cohort-day Cox models, and genus- or
subfamily-level contrasts resting on few tip taxa are interpreted as
provisional.

Our approach is survival modelling --- estimating how
instantaneous mortality hazard changes as a function of candidate predictors
--- not bivariate association.  Cox proportional-hazards models provide
directional, time-to-event inference with censoring, clustering, and
simultaneous covariates; where we report correlations (e.g.\ Spearman
$\rho$), they serve as auxiliary checks, not as the primary evidence.  The
inferential strategy rests on \emph{mechanistic exclusion}: by testing three
candidate moderators of the size--longevity effect --- finding colony size
and temperature null, and a weak but significant foraging-rate interaction ---
we narrow the mechanistic window rather than simply documenting a
relationship.  Finally, we test for nonlinearity in the temperature--hazard
function (piecewise spline with Davies permutation) and for effect
heterogeneity across subfamilies (interaction terms), both of which go beyond
what a correlational framework can address.

%% ---- Analytical roadmap (Box 1 -> Table 1 in manuscript) -----
\begin{table}[htbp]
\caption{Analytical roadmap: questions, claims, internal hypotheses, and
  outcomes.  Internal labels (H1--H9) cross-reference the methods, code, and
  supplementary tables.
  \label{tab:roadmap}}
\begin{threeparttable}
\small
\begin{tabularx}{\textwidth}{X r}
\toprule
\textbf{Analysis} & \textbf{Outcome}\\
\midrule
\multicolumn{2}{l}{\textbf{Q1.\ What mechanism links body size to
  longevity?}}\\
\multicolumn{2}{l}{\emph{Claim~1.\ Body size predicts duration; intrinsic
  physiology inferred by exclusion.}}\\[4pt]
\quad Larger workers live longer (H1)
  & $\checkmark$ HR\,=\,0.67, $p = 0.002$\\
\quad Colony size does not modulate it (H2a)
  & $\circ$ $p = 0.60$\\
\quad Foraging rate modulates it weakly (H2b)
  & $\checkmark$ LRT $p = 0.014$ (M2)\\
\quad Body size does not buffer heat (H3)
  & $\circ$ HR\,=\,1.00, $p = 0.72$\\
\quad Size effect stronger in field (H2c)
  & $\triangleright$ 1.28$\times$\\
\midrule
\multicolumn{2}{l}{\textbf{Q2.\ What shapes the trajectory of
  mortality?}}\\
\multicolumn{2}{l}{\emph{Claim~2.\ Trajectory associated with circadian
  niche, not body size.}}\\[4pt]
\quad Senescence is the rule (H4)
  & $\checkmark$ 14/18 $\rho > 1$\\
\quad Trajectory is associated with circadian niche (H5)
  & $\checkmark$ KW $p = 0.009$\\
\quad Field steepens senescence (H6)
  & $\checkmark$ 17/18, $p < 0.001$\\
\quad Trajectory unrelated to body mass
  & $\circ$ Sp.\ $\rho = -0.25$, $p = 0.32$\\
\midrule
\multicolumn{2}{l}{\textbf{Q3.\ How does temperature act?  Is risk
  taxon-specific?}}\\
\multicolumn{2}{l}{\emph{Claim~3.\ Thermal risk nonlinear and
  taxonomically patterned.}}\\[4pt]
\quad Temperature raises hazard (H7)
  & $\checkmark$ HR\,=\,1.036, $p < 0.001$\\
\quad Hazard plateaus above 20\,\textdegree C (H8)
  & $\checkmark$ $T^* = 20$\,\textdegree C, $p < 0.001$\\
\quad Vulnerability is subfamily-specific (H9)
  & $\checkmark$ joint $p = 0.017$\\
\bottomrule
\end{tabularx}
\begin{tablenotes}
\footnotesize
\item[$\checkmark$] Supported. \quad $\circ$ Null result. \quad
  $\triangleright$ Directional (consistent, not conclusive).
\item \textbf{Abbreviations:} HR, hazard ratio (Cox proportional-hazards
  model); KW, Kruskal--Wallis test; Sp., Spearman rank correlation;
  $T^*$, estimated temperature breakpoint; $\rho$, Weibull shape parameter.
  Fractions of the form $X/18$ refer to the 18 study species.  Internal
  hypothesis labels (H1--H9) are defined in the Methods
  (\S\ref{sec:cox}--\S\ref{sec:thermal}).
\end{tablenotes}
\end{threeparttable}
\end{table}

\subsection{Study system and data}
\label{sec:data}

We re-analyse the field mark--recapture dataset of \citet{Riskas2026},
comprising 18 Australian ant species monitored at the La~Trobe Wildlife
Sanctuary, a 28-ha reserve on La~Trobe University's Melbourne campus,
Victoria, in temperate south-eastern Australia (${\sim}37.7^\circ$\,S,
$145.1^\circ$\,E).  The site has a mesic, temperate climate (mean annual
minimum and maximum air temperatures of 9.7\,\textdegree C and
20.2\,\textdegree C; ${\sim}656$\,mm annual rainfall;
\citealt{BoMBundoora}).
The field dataset contains 2,363 cohort-day observations spanning 106
cohorts across 39 colonies, collected between February 2022 and May 2023,
with 1,826 recorded daily mortality events.  Daily maximum temperature
(\texttt{maxt}) averaged 22.8\,\textdegree C (range 13.1--40.5\,\textdegree C).
For laboratory comparisons (Claims 1 and 2) we use the paired laboratory
dataset of 1,148 individual workers (39 colonies, 18 species) maintained
under controlled temperature and humidity.  Full field-collection protocol
and laboratory rearing conditions are described in \citet{Riskas2026}.
Descriptive statistics for all 18 species are given in
Table~\ref{tab:descriptive}.

Species were chosen to span the full range of body sizes, subfamily
identities, and circadian activity regimes present in this temperate
assemblage, following the sampling design of \citet{Riskas2026}.  Colonies
were recruited opportunistically at the La~Trobe Wildlife Sanctuary by active
search and pitfall trapping, with inclusion conditional on at least two
successfully marked cohorts per colony.  The resulting sample --- 18 species
across five subfamilies --- captures enough interspecific variation to
estimate Cox and Weibull parameters, but at $N = 18$ species it is too small
to support fully factorial inference across all three mortality axes
simultaneously; we return to this constraint in the Limitations
(\S\ref{sec:limitations}).

%% Fig 1 — Study organism (FE convention: organism/study area first)
\begin{figure}[htbp]
\centering
\includegraphics[width=0.65\textwidth]{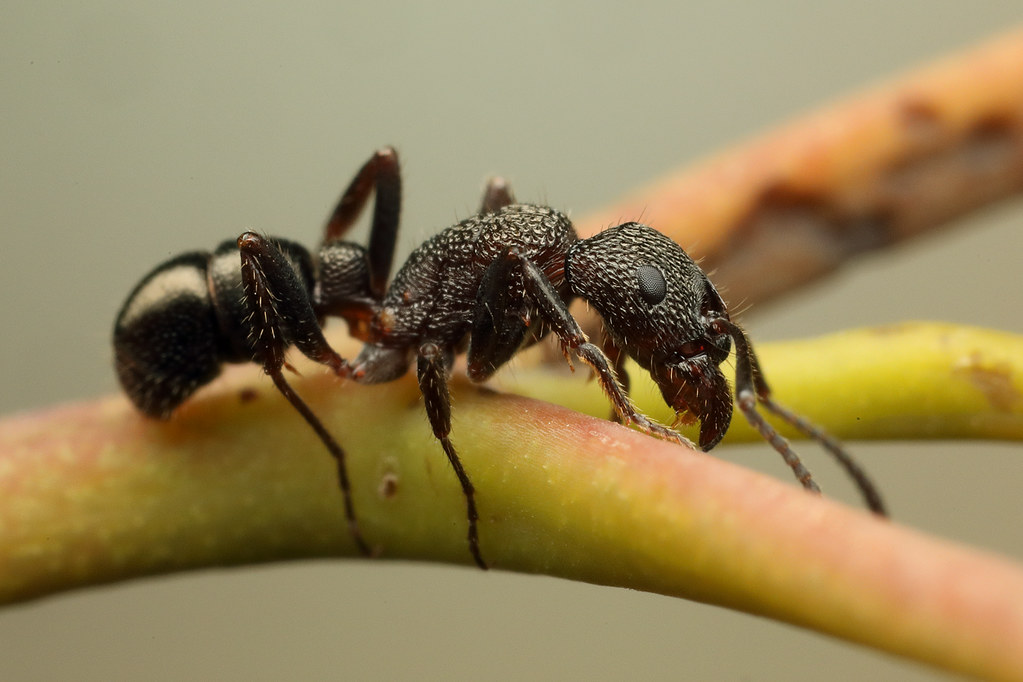}
\caption{\textbf{Representative study organism: \textit{Rhytidoponera}
  sp.\ (Ectatomminae).}  A worker of \textit{Rhytidoponera} sp., one of
  three ectatomminine species included in the field mark--recapture dataset
  (Ectatomminae: \textit{R.~metallica}, \textit{R.~tasmaniensis},
  \textit{R.~victoriae}).  This subfamily is distinguished by notably low
  colony size (mean\,$\approx$170 workers) relative to myrmicine and
  dolichoderine species, the steepest actuarial senescence trajectories in
  laboratory survival assays ($\rho_{\rm lab} \approx 1.4$--$2.0$), and
  elevated thermal hazard sensitivity in the field.
  Photo: \textit{Rhytidoponera} sp.\ by P.\,Kavanagh
  (\protect\url{https://www.flickr.com/photos/patrick_k59/49413794622}),
  CC~BY~2.0.
  \label{fig:organism}}
\end{figure}

%% ---- Table 2: Descriptive statistics (18 species) -----------
\begin{sidewaystable}
\caption{Species-level descriptive statistics for the 18 study species,
  grouped by subfamily and sorted by log\textsubscript{10} body mass within
  subfamily.  Columns: \textbf{log\textsubscript{10} mass (mg)}, mean worker
  wet mass; \textbf{Colony size}, mean number of workers per colony;
  \textbf{Cohorts ($N$)}, number of field mark--recapture cohorts;
  \textbf{Median obs (d)}, median per-worker observation window in days;
  \textbf{Foraging rate}, mean workers leaving the nest per minute;
  \textbf{\texttt{maxt} mean (\textdegree C)}, mean daily maximum air
  temperature experienced; $\boldsymbol{\rho_{\rm lab}}$, Weibull shape
  parameter from individual laboratory survival; $\boldsymbol{\rho_{\rm
  field}}$, Weibull shape parameter from field cohort survival.  For all
  $\rho$, values $> 1$ indicate senescence, $\approx 1$ constant hazard, and
  $< 1$ frailty selection.  Morphospecies designators (e.g.\ \textit{Camponotus
  claripes}~A, \textit{Papyrius}~A) follow the genus name in roman type.
  \label{tab:descriptive}}
\begin{threeparttable}
\footnotesize
\begin{tabular}{llrrrrrrrr}
\toprule
\textbf{Species} & \textbf{Subfamily} &
\multicolumn{1}{c}{\textbf{log\textsubscript{10}}} &
\multicolumn{1}{c}{\textbf{Colony}} &
\multicolumn{1}{c}{\textbf{Cohorts}} &
\multicolumn{1}{c}{\textbf{Median}} &
\multicolumn{1}{c}{\textbf{Foraging}} &
\multicolumn{1}{c}{\textbf{\texttt{maxt}}} &
\multicolumn{1}{c}{$\boldsymbol{\rho_{\rm lab}}$} &
\multicolumn{1}{c}{$\boldsymbol{\rho_{\rm field}}$}\\
 & &
\textbf{mass (mg)} & \textbf{size} & $(N)$ & \textbf{obs (d)} &
\textbf{rate} & \textbf{mean (\textdegree C)} & & \\
\midrule
\textit{Rhytidoponera metallica}    & Ectatomminae  & 0.66 & 61   &  7 & 81.0 & 0.74 & 23.3 & 1.38 & 0.54\\
\textit{Rhytidoponera tasmaniensis} & Ectatomminae  & 0.37 & 194  &  8 & 44.0 & 0.74 & 22.4 & 1.99 & 2.94\\
\textit{Rhytidoponera victoriae}    & Ectatomminae  & $-$0.18 & 257  &  6 & 76.5 & 0.93 & 26.0 & 1.70 & 3.59\\
\midrule
\textit{Camponotus consobrinus}     & Formicinae    & 0.94 & 192  &  8 & 94.5 & 0.86 & 24.4 & 1.08 & 2.85\\
\textit{Camponotus claripes}~A      & Formicinae    & 0.33 & 161  &  4 & 64.5 & 0.73 & 23.5 & 1.15 & 4.89\\
\textit{Camponotus nigroaeneus}     & Formicinae    & $-$0.09 & 295  &  2 & 61.0 & 0.70 & 23.7 & 0.89 & 5.65\tnote{$\dagger$}\\
\textit{Melophorus turneri}         & Formicinae    & $-$0.81 & 871  &  8 & 61.0 & 1.81 & 26.7 & 0.96 & 5.15\\
\midrule
\textit{Myrmecia pyriformis}        & Myrmeciinae   & 1.21 & 127  &  6 & 121.5 & 0.64 & 24.6 & 0.83 & 1.25\\
\textit{Myrmecia piliventris}       & Myrmeciinae   & 0.98 & 150  &  3 & 91.0 & 0.66 & 20.6 & 0.81 & 4.47\\
\midrule
\textit{Meranoplus fenestratus}     & Myrmicinae    & 0.09 & 179  &  8 & 96.0 & 0.87 & 24.8 & 1.85 & 3.28\\
\textit{Crematogaster laeviceps}    & Myrmicinae    & $-$0.45 & 2123 &  2 & 75.5 & 2.64 & 25.9 & 1.82 & 5.65\tnote{$\dagger$}\\
\textit{Chelaner cinctum}           & Myrmicinae    & $-$0.68 & 472  & 10 & 83.5 & 1.06 & 25.1 & 1.46 & 1.34\\
\textit{Meranoplus diversus}        & Myrmicinae    & $-$0.71 & 142  &  5 & 123.0 & 0.84 & 26.1 & 1.47 & 2.60\\
\midrule
\textit{Anonychomyrma nitidiceps}   & Dolichoderinae & $-$0.33 & 16666 & 6 & 89.5 & 7.25 & 22.3 & 1.56 & 2.12\\
\textit{Iridomyrmex septentrionalis} & Dolichoderinae & $-$0.59 & 7964 & 8 & 50.0 & 8.54 & 25.3 & 1.07 & 4.69\\
\textit{Iridomyrmex splendens}      & Dolichoderinae & $-$0.61 & 3155 & 5 & 57.0 & 2.10 & 25.2 & 1.12 & 3.99\\
\textit{Papyrius} A                 & Dolichoderinae & $-$0.89 & 5711 & 8 & 73.0 & 5.58 & 25.9 & 1.01 & 3.79\\
\textit{Iridomyrmex notialis}       & Dolichoderinae & $-$0.95 & 1006 & 2 & 70.0 & 1.58 & 21.2 & 2.51 & 5.65\tnote{$\dagger$}\\
\bottomrule
\end{tabular}
\begin{tablenotes}
\footnotesize
\item[$\dagger$] Field $\rho$ estimated from pooled assemblage AFT fallback
  (fewer than 3 cohorts available for separate Weibull fit).
\end{tablenotes}
\end{threeparttable}
\end{sidewaystable}

\subsubsection*{Replication Statement}

\begin{table}[h!]
\centering
\caption*{\textit{Replication Statement} (Functional Ecology requirement)}
\small
\begin{tabularx}{\linewidth}{lXl}
\toprule
\textbf{Scale of inference} & \textbf{Factor(s) applied} &
\textbf{Number of replicates}\\
\midrule
Species ($N = 18$) &
Body mass (log\textsubscript{10}, centred); circadian niche (3 groups);
subfamily (5 groups) &
18 species\\
Cohort &
Daily mortality event; observation day &
106 cohorts, 39 colonies\\
Assemblage &
Daily maximum air temperature (\texttt{maxt}, \textdegree C) &
2,363 cohort-day observations\\
Individual (lab) &
Body mass; colony identity &
1,148 workers, 39 colonies\\
\bottomrule
\end{tabularx}
\end{table}

Each of the three claims maps to a single pre-specified primary endpoint
(Table~\ref{tab:roadmap}): the Cox M0 body-size hazard ratio for duration
(H1); the laboratory Weibull shape $\rho_{\rm lab}$, tested across circadian
niches by Kruskal--Wallis, for senescence trajectory (H5); and a piecewise
Cox model selected by AIC, with breakpoint significance from a Davies
permutation test, for the thermal plateau (H8).  We control confounding
through three complementary devices rather than a single adjustment:
cluster-robust standard errors (cluster\,=\,cohort) absorb within-cohort
autocorrelation; subfamily fixed effects absorb shared phylogenetic baseline
hazard; and a suite of sensitivity analyses (species fixed effects,
stratified Cox, and exclusion of pooled-Weibull fallback species) probes
robustness to the strongest assumptions.  Because the three claims address
orthogonal null hypotheses by design, we do not apply a multiple-comparison
correction across them.

\subsection{Duration models --- Cox proportional hazards (Claim 1)}
\label{sec:cox}

We fit Cox proportional-hazards (PH) models to the cohort-day dataset with
cluster-robust standard errors (cluster\,=\,\texttt{cohort\_id}) to account
for within-cohort temporal autocorrelation.  The event is the daily mortality
indicator (\texttt{event\_wd}), the time variable is days since cohort
formation (\texttt{Day}), and the focal predictor is log\textsubscript{10}
worker body mass (\texttt{logw\_mass}, centred as \texttt{logw\_c}).
Subfamily (\texttt{SF}) is a fixed effect with Dolichoderinae as reference.
The proportional hazards assumption was verified using Schoenfeld residuals
(\texttt{lifelines} \texttt{proportional\_hazard\_test}, rank transform):
no covariate showed a significant time-trend ($p \geq 0.17$ for all
predictors including body size and \texttt{maxt}; Table~\ref{tab:schoenfeld}
in supplementary materials).  As a species-frailty robustness check
(supplementary Table~\ref{tab:frailty}), we re-fitted model M0 with (i)
species fixed effects (\texttt{C(SPEC)}, 18 dummies) and (ii) stratification
by species (\texttt{strata = [SPEC]}).  Both variants yielded a
directionally consistent body-size coefficient (HR\,=\,0.77--0.71,
respectively), though non-significant within species where intraspecific
size variation is narrow; the hazard ratio obtained from M0 therefore
reflects a predominantly interspecific size--longevity relationship.
To test whether genus-level clustering further confounds the result
(several genera contribute two or three species: \textit{Rhytidoponera},
\textit{Iridomyrmex}, \textit{Camponotus}, \textit{Meranoplus}), we
attempted to fit M0 stratified by genus.  The model failed to converge
due to matrix singularity, indicating that genus and subfamily structures
are too collinear to separate at $N = 18$; the subfamily fixed-effect
parameterisation of M0 is therefore retained.

The analysis variables are derived from the raw mark--recapture records as
follows.  The cohort-day table has one row per cohort-day combination, with
the event set to \texttt{event\_wd} (1 if at least one worker died on that
calendar day, 0 otherwise) and time measured as days since cohort initiation
(\texttt{Day}).  Body mass (\texttt{logw\_mass}) is the
log\textsubscript{10} of worker wet mass in mg, centred at the assemblage
mean (\texttt{logw\_c}\,$=\,$\texttt{logw\_mass}\,$-\,\bar{x}$).  Daily
maximum temperature (\texttt{maxt}) is taken from the nearest Bureau of
Meteorology station (Bundoora, ${\sim}1$\,km from the site) and, for
interaction models, centred analogously (\texttt{maxt\_c}).  The three
species with fewer than three field cohorts (\textit{Camponotus
nigroaeneus}, \textit{Crematogaster laeviceps}, \textit{Iridomyrmex
notialis}) receive $\rho_{\rm field}$ from a pooled assemblage Weibull AFT
model and are flagged $\dagger$ throughout.

\begin{itemize}
  \item \textbf{H1} (size $\to$ longevity): \emph{Do larger workers survive
    longer?}  Expected: HR\,$< 1$ for \texttt{logw\_c}.  Model:
    \texttt{hazard\,$\sim$\,logw\_c + maxt + C(SF)} (M0).

  \item \textbf{H2a} (colony-size moderator): \emph{Does colony size
    weaken the per-worker size benefit (colony-dilution hypothesis)?}
    Expected: null interaction if size acts intrinsically.  OLS on cohort
    means ($N = 39$ colonies):
    \texttt{total\_days\,$\sim$\,logw\_c\,$\times$\,cs\_c + C(SF)}.

  \item \textbf{H2b} (foraging-rate moderator): \emph{Does foraging intensity
    modulate size-longevity?}  Expected: null if size acts primarily via
    intrinsic physiology.
    Model M2 (the size\,$\times$\,foraging model):
    \texttt{hazard\,$\sim$\,logw\_c\,$\times$\,fr\_c + maxt\_c + C(SF)}
    (Table~\ref{tab:cox_compare}).

  \item \textbf{H2c} (field amplification of size): \emph{Is the size
    benefit stronger in the field than in the lab?}  Compares
    $|\hat\beta_\text{logw}|$ between field and laboratory Cox models, plus
    a species-level Spearman correlation of survival indices ($\Phi$;
    Fig.~\ref{fig:forest}).

  \item \textbf{H3} (thermal-buffer null): \emph{Does body size buffer
    thermal mortality (extrinsic-buffering hypothesis)?}  Expected: null
    interaction if size acts intrinsically.  Model M1 (the size\,$\times$\,temperature model):
    \texttt{hazard\,$\sim$\,logw\_c\,$\times$\,maxt\_c + C(SF)}.
\end{itemize}

Nested models are compared by partial likelihood-ratio test (LRT) and AIC.

\subsection{Trajectory models --- Weibull survival (Claim 2)}
\label{sec:weibull}

We fit Weibull survival models per species to estimate the shape parameter
$\rho$ (senescence rate) and scale $\lambda$.  Laboratory $\rho_{\rm lab}$
is estimated from individual-level data (one row per worker; event = death;
time = days to event or right-censoring) using maximum-likelihood Weibull
regression per species.  Field $\rho_{\rm field}$ is estimated from the
cohort-day counting-process data (one row per worker-day; event = daily
mortality indicator; time window = entry/exit day for that worker-day
record).  For each species with three or more cohorts, a separate Weibull
proportional-hazards model is fitted to that species' cohort-day records,
extracting $\rho_{\rm field}$ as the estimated shape.  For three species
with fewer than three cohorts (\textit{Camponotus nigroaeneus},
\textit{Crematogaster laeviceps}, \textit{Iridomyrmex notialis}), a pooled
assemblage AFT model supplies a fallback $\rho_{\rm field}$ estimate;
these fallback cases are flagged with $\dagger$ in all tables and figures.
To quantify the collinearity between circadian niche and subfamily as
explanatory variables of $\rho_{\rm lab}$, we performed a variance
partitioning analysis~\citep{Legendre2012}: three OLS models
(niche-only, subfamily-only, and combined) were fitted on the 18
species-level $\rho_{\rm lab}$ estimates, and the unique and shared
fractions of adjusted $R^2$ were computed following Legendre and
Legendre (2012).

\begin{itemize}
  \item \textbf{H4} (senescence is prevalent): \emph{Is actuarial senescence
    ($\rho > 1$) the dominant life-history pattern, rather than frailty
    selection ($\rho < 1$)?}  Assessed as the proportion of species with
    $\rho > 1$ and 95\,\% CI above 1.

  \item \textbf{H5} (circadian structuring of trajectory): \emph{Is
    senescence rate associated with circadian activity niche rather than body size?}
    $\rho$ compared among niches (matinal/diurnal/crepuscular) by
    Kruskal--Wallis with pairwise follow-up (Dunn's test with
    Bonferroni correction, implemented via \texttt{scikit-posthocs}
    \texttt{posthoc\_dunn}), and correlated with body mass
    (Spearman).  Circadian niche assigned from peak foraging hour
    (\texttt{peak\_niche}; \citealt{Riskas2026}).

  \item \textbf{H6} (field amplification of trajectory): \emph{Do field
    conditions steepen senescence trajectories relative to the laboratory?}
    $\Delta\rho = \rho_{\rm field} - \rho_{\rm lab}$ tested against zero
    by Wilcoxon signed-rank ($\Delta\rho > 0$ = steeper field senescence;
    Table~\ref{tab:delta_rho}).
\end{itemize}

\subsection{Thermal-risk models --- Cox PH and piecewise spline (Claim 3)}
\label{sec:thermal}

We model the assemblage-level temperature--hazard relationship, test it for
nonlinearity, and ask whether thermal vulnerability differs among subfamilies.

\begin{itemize}
  \item \textbf{H7} (temperature main effect): \emph{Does higher daily
    maximum temperature increase mortality hazard at the assemblage level?}
    Expected: HR\,$> 1$ for \texttt{maxt}.  Model M0:
    \texttt{hazard\,$\sim$\,logw\_c + maxt + C(SF)}; interest is
    $\hat\beta_\text{maxt}$ (HR per \textdegree C).  H7 establishes the
    assemblage-level baseline for H8 and H9.

  \item \textbf{H8} (nonlinear plateau): \emph{Is the temperature--hazard
    relationship nonlinear, with a plateau above some breakpoint $T^*$
    consistent with behavioural buffering?}  Piecewise-linear spline
    models are fitted across $T^* \in \{15, \ldots, 34\}$\,\textdegree C:
    \texttt{hazard\,$\sim$\,logw\_c + maxt + (maxt\,$-$\,$T^*$)$_+$ + C(SF)};
    best $T^*$ by AIC; significance by Davies permutation test
    (Table~\ref{tab:spline}).

  \item \textbf{H9} (taxon-specific vulnerability): \emph{Do subfamilies
    differ in thermal sensitivity, predicting that some lineages remain
    exposed above the plateau?}  Subfamily $\times$ temperature interactions
    added:
    \texttt{hazard\,$\sim$\,logw\_c + maxt\_c + $\Sigma_i$(maxt\_c\,$\times$\,SF$_i$)
    + C(SF)};  reference category for all subfamily contrasts is
    Dolichoderinae (the most species-rich group), encoded as the omitted
    dummy in \texttt{C(SF)}.  Significance assessed by joint LRT
    ($\chi^2(4)$) first; individual post-hoc comparisons (each
    SF$_i$\,$\times$\,\texttt{maxt\_c} Wald test) reported only upon rejection
    of the joint null.  A binary variant uses a heat-event indicator
    (\texttt{heat30}, maxt\,$> 30$\,\textdegree C;
    Table~\ref{tab:subfamily_temp}).
\end{itemize}

\subsection{Software}
\label{sec:software}

All analyses use Python~3.13: Cox~PH and Weibull models with
\texttt{lifelines}~v0.29; OLS/WLS with \texttt{statsmodels}~v0.14; numerical
routines with \texttt{numpy}/\texttt{scipy}; figures with
\texttt{matplotlib}~v3.9.  A fixed random seed (42) is used throughout.  All
code and processed data are available at the Open Science Framework
(\url{https://osf.io/[FILL-BEFORE-SUBMISSION]}): the Python pipeline,
processed CSVs, and a \texttt{README} with step-by-step reproduction
instructions.

%% ============================================================
\section{Results}
\label{sec:results}

\subsection{Claim 1 --- Body size predicts duration; colony size and temperature
  are null moderators, with a weak foraging-rate interaction}
\label{sec:claim1}

\begin{figure}[htbp]
\centering
\includegraphics[width=0.85\textwidth]{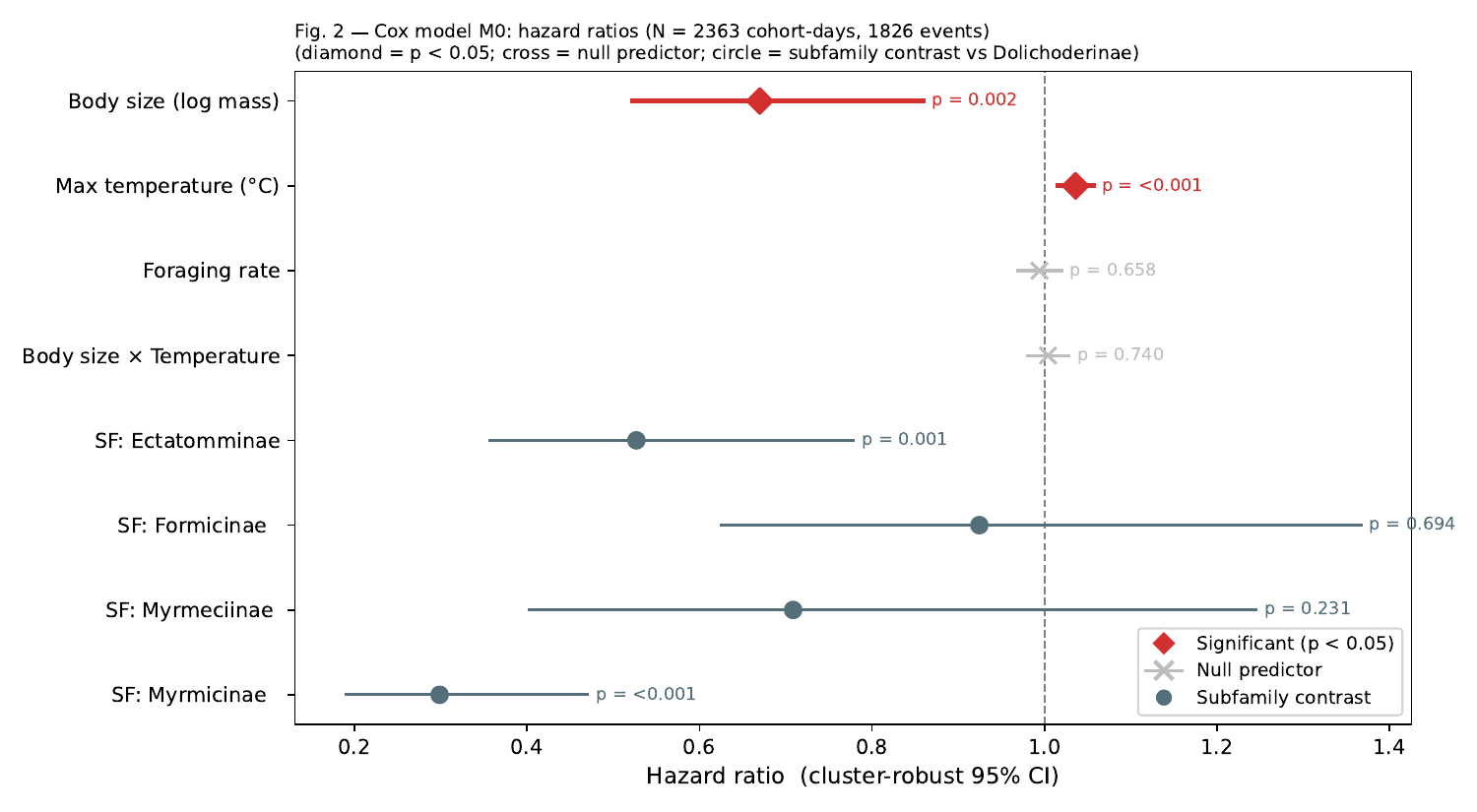}
\caption{\textbf{Forest plot of hazard ratios from the main Cox
  proportional-hazards model M0 (Claims~1 and~3).}  Each row shows a
  predictor with its point estimate and cluster-robust 95\,\% CI
  (cluster\,=\,\texttt{cohort\_id}).  Marker shape encodes effect type
  independently of colour: \emph{diamonds} mark statistically significant
  predictors (Wald $p < 0.05$: body size and maximum temperature);
  \emph{crosses} mark null predictors (body
  size\,$\times$\,temperature interaction); \emph{circles} mark subfamily
  baseline-hazard contrasts relative to the Dolichoderinae reference.
  Vertical dashed reference line at HR\,=\,1.  Model fitted on
  $N = 2{,}363$ cohort-day observations (1{,}826 events).
  M0, primary Cox model (\texttt{logw\_c + maxt + C(SF)}); HR, hazard ratio;
  CI, confidence interval; SF, subfamily.
  \label{fig:forest}}
\end{figure}

Body size has a strong, largely intrinsic (context-independent) effect
on longevity that is insensitive to colony size and thermal context.  In Cox
proportional-hazards model M0 (\texttt{logw\_c + maxt + C(SF)},
cluster-robust, $N = 2{,}363$ cohort-days), each log\textsubscript{10} unit
increase in body mass reduces daily hazard by 33\,\% (HR\,=\,0.67, 95\,\%\,CI
[0.52--0.87], $p = 0.002$; Table~\ref{tab:cox_main}), consistent with the
foundational size--longevity relationship established in \citet{Riskas2026}.
The main-effects model has
moderate discrimination (concordance\,=\,0.637; AIC\,=\,24{,}542), and among
subfamilies Myrmicinae (HR\,=\,0.31, $p < 0.001$) and Ectatomminae
(HR\,=\,0.55, $p = 0.004$) carry markedly lower baseline hazard than the
Dolichoderinae reference (Fig.~\ref{fig:forest}).

The size--longevity relationship is a between-species signal.  When species
identity was absorbed as fixed effects (18 dummies) or as Cox strata, the
body-size coefficient remained directionally consistent but became
non-significant (HR\,=\,0.77--0.71, $p = 0.47$--$0.35$;
Table~\ref{tab:frailty}), reflecting narrow intraspecific body-size
variation rather than an absence of effect.  The HR\,=\,0.67 reported
throughout therefore quantifies a between-species comparative pattern ---
a distinction that matters for applying size-based vulnerability indices
to species comparisons versus within-colony trait variation.

Phylogenetic generalised least squares (PGLS) confirmed that this
size--longevity signal carries negligible phylogenetic structure: Pagel's
$\hat\lambda = 0.06$ (LRT $\chi^2(1) = 0.02$, $p = 0.90$), indicating
that species-level body-size log-hazard estimates show no detectable
covariation with shared evolutionary history at this phylogenetic depth.
The non-phylogenetic Cox model (M0) is therefore retained as the primary
model (Supplementary Table~\ref{tab:pgls_size}).

%% Fig 3 — Forest plot

%% ---- Table 4: Main Cox coefficients (M0) --------------------
\begin{table}[htbp]
\caption{Main Cox proportional-hazards model (M0) coefficients (Claims~1
  and~3, H1, H3, H7).  Reference category: Dolichoderinae.  HR, hazard ratio;
  CI, 95\,\% confidence interval; $z$, Wald test statistic.
  \label{tab:cox_main}}
\begin{threeparttable}
\begin{tabular}{lrrrrl}
\toprule
\textbf{Covariate} & $\boldsymbol{\beta}$ & \textbf{SE} & \textbf{HR} &
\textbf{95\,\% CI} & \textbf{\textit{p}}\\
\midrule
Body size (log\textsubscript{10} mass, centred) &
  $-$0.398 & 0.130 & 0.67 & [0.52, 0.87] & 0.002\\
Max.\ temperature (\textdegree C) &
  $+$0.035 & 0.010 & 1.036 & [1.016, 1.057] & $< 0.001$\\
\midrule
Subfamily: Ectatomminae &
  $-$0.600 & 0.210 & 0.55 & [0.36, 0.83] & 0.004\\
Subfamily: Formicinae &
  $-$0.105 & 0.193 & 0.90 & [0.62, 1.32] & 0.589\\
Subfamily: Myrmeciinae &
  $-$0.399 & 0.280 & 0.67 & [0.39, 1.16] & 0.155\\
Subfamily: Myrmicinae &
  $-$1.205 & 0.231 & 0.31 & [0.19, 0.47] & $< 0.001$\\
\midrule
\multicolumn{6}{l}{\textit{Size\,$\times$\,temperature interaction (H3):
  HR\,=\,1.00 [0.98, 1.03], $p = 0.72$; $\Delta$AIC\,=\,$+$1.45 vs.\ M0}}\\ 
\bottomrule
\end{tabular}
\begin{tablenotes}
\footnotesize
\item Concordance\,=\,0.64; AIC\,=\,24{,}542; $N = 2{,}363$ cohort-day obs.,
  106 cohorts, 39 colonies.  Cluster-robust SEs (cluster\,=\,cohort).
\end{tablenotes}
\end{threeparttable}
\end{table}

\begin{figure}[htbp]
\centering
\includegraphics[width=0.85\textwidth]{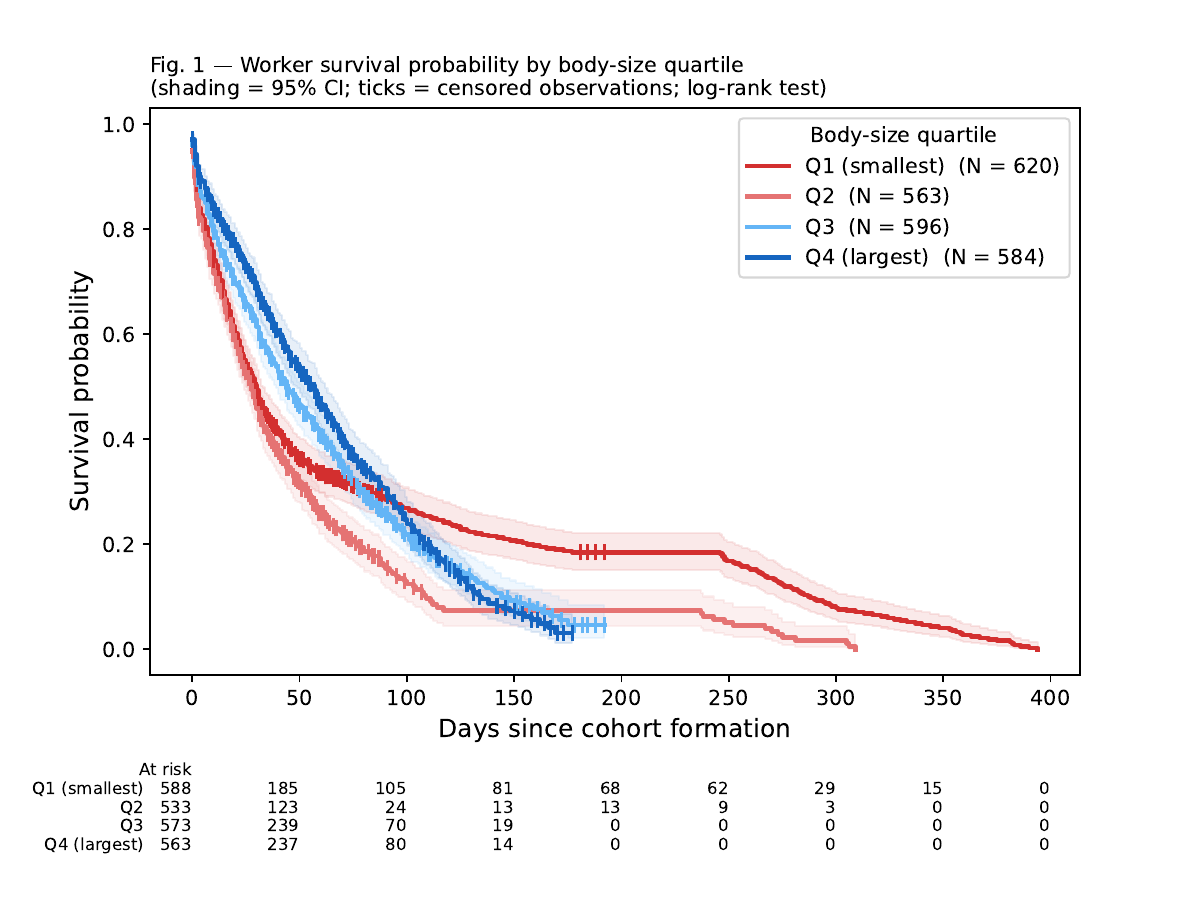}
\caption{\textbf{Kaplan--Meier survival curves by body-size quartile
  (Claim~1, H1).}  Field workers are grouped into four body-size quartiles
  (Q1\,=\,smallest, $N = 620$; Q2, $N = 563$; Q3, $N = 596$;
  Q4\,=\,largest, $N = 584$; quartile boundaries defined from the full
  cohort-day field dataset).  Shaded bands are pointwise 95\,\% confidence
  intervals; vertical ticks mark censored observations; the number of
  individuals at risk is tabulated below the time axis.  The stepwise
  separation between quartiles illustrates the intrinsic size--longevity
  relationship: larger workers survive significantly longer across the full
  observation window (log-rank $p < 0.001$; Cox HR per log\textsubscript{10}
  unit\,=\,0.67, 95\,\%\,CI [0.52--0.87], model M0,
  Table~\ref{tab:cox_main}).
  \label{fig:km}}
\end{figure}

\begin{figure}[htbp]
\centering
\includegraphics[width=0.95\textwidth]{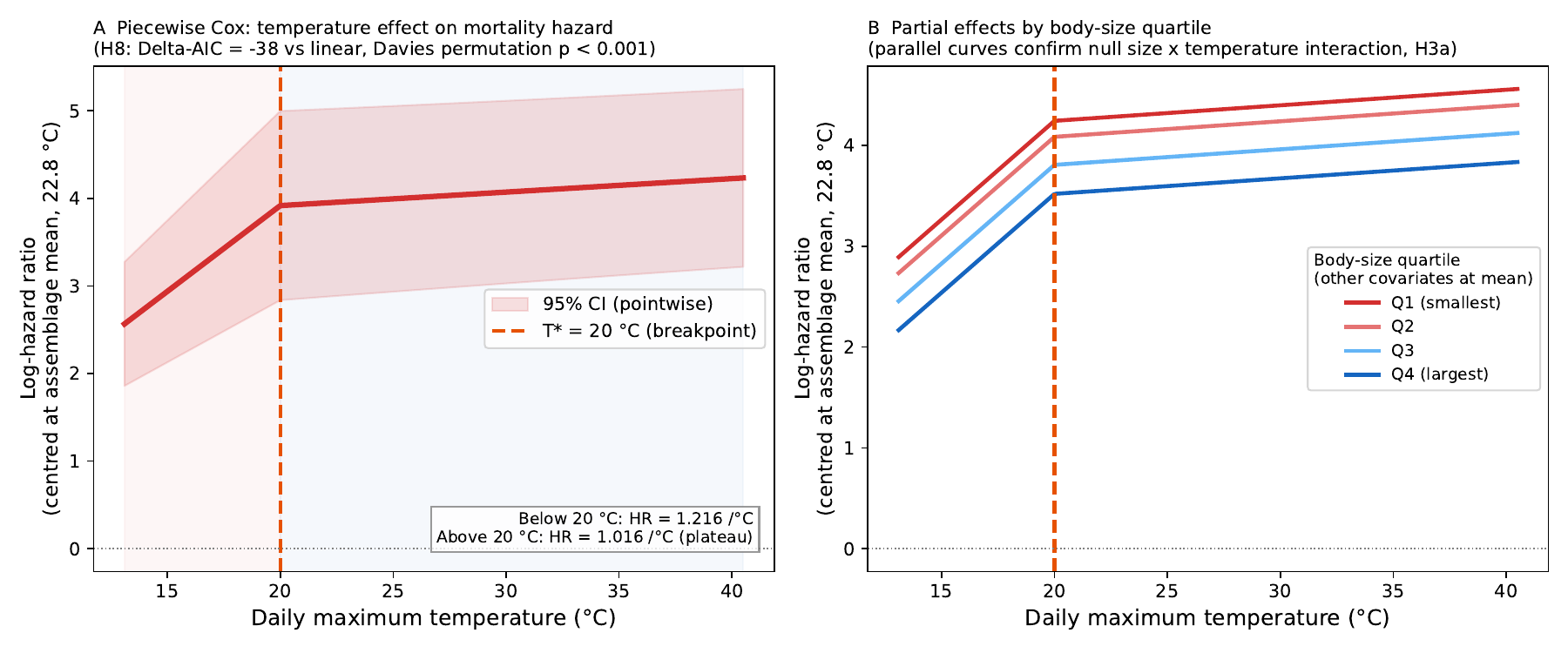}
\caption{\textbf{Temperature--mortality relationship under the piecewise
  spline model (Claim~3, H8).}  \textbf{Panel~A:} Log-hazard ratio (centred
  at the assemblage mean daily maximum temperature,
  $\bar{T}_{\max} = 22.8$\,\textdegree C) as a function of daily maximum
  temperature (\texttt{maxt}), with the estimated breakpoint
  $T^* = 20$\,\textdegree C marked by a dashed vertical line.  The shaded
  band is the pointwise 95\,\% confidence interval derived from the
  cluster-robust variance matrix of the piecewise Cox model; faint
  background shading distinguishes the rising segment (below
  20\,\textdegree C; HR\,=\,1.216/\textdegree C) from the near-plateau
  segment (above 20\,\textdegree C; HR\,=\,1.016/\textdegree C).  The
  piecewise model fits substantially better than a linear model
  ($\Delta$AIC\,=\,$-38$; Davies permutation $p < 0.001$;
  Table~\ref{tab:spline}).  \textbf{Panel~B:} Partial-effect hazard curves
  for each body-size quartile over the same temperature range, with all
  other covariates held at their means.  The four curves are parallel,
  consistent with the null size\,$\times$\,temperature interaction
  (H3, Table~\ref{tab:cox_main}), and the
  plateau is equally present across all size classes.
  \label{fig:spline}}
\end{figure}

Of the three contexts we tested, two did not moderate the size effect and
one showed a weak interaction.  First,
colony size: the body-size\,$\times$\,colony-size interaction was null
(OLS interaction $p = 0.60$, 95\,\%\,CI [$-$58, $+$77]\,days, $N = 39$
colonies; Table~\ref{tab:ols}).  Second, foraging rate: the
body-size\,$\times$\,foraging-rate interaction was weak but significant
(M2 vs.\ M0$_{\rm fr}$: LRT $\chi^2(1) = 6.08$, $p = 0.014$,
$\Delta$AIC\,=\,$-$4.1; Table~\ref{tab:cox_compare}).  Thus only colony size
and temperature remain clearly null moderators; the foraging interaction
qualifies the otherwise intrinsic reading and should be interpreted
cautiously given moderate collinearity (VIF\,$\approx$\,4;
\S\ref{sec:discussion}).
Third, temperature: in Cox model M1\textsubscript{temp} (M0 extended with the
body size\,$\times$\,temperature product), the interaction term was
essentially absent (HR\,=\,1.00, 95\,\%\,CI [0.98--1.03], $p = 0.72$;
$\Delta$AIC\,=\,$+$1.45, LRT $p = 0.46$ vs.\ M0; Fig.~\ref{fig:spline}B), and
hazard curves across body-size quartiles remained parallel as temperature
rose (Fig.~\ref{fig:km}).  The interaction CI [0.98, 1.03] corresponds to an
upper bound of $\approx$\,3\,\% per \textdegree C on any thermal moderation of
the size benefit, against a main temperature effect of HR\,=\,1.04
($\approx$\,3.6\,\%/\textdegree C).

%% Fig 2 — KM curves

%% Fig 4 — Spline/temperature

%% ---- Table 5: Colony-level OLS (H2a) ------------------------
\begin{table}[htbp]
\caption{Colony-level OLS regression of total lifespan on body size and
  colony size (H2a).  Interest is the body-size\,$\times$\,colony-size
  interaction (null hypothesis: colony dilution does not moderate the size
  benefit).  Reference: Dolichoderinae.
  \label{tab:ols}}
\begin{threeparttable}
\begin{tabular}{lrrrrl}
\toprule
\textbf{Term} & $\boldsymbol{\beta}$ & \textbf{SE} & $\boldsymbol{t}$ &
\textbf{95\,\% CI} & \textbf{\textit{p}}\\
\midrule
Intercept & 71.9 & 35.5 & 2.02 & [$-$1.0,  144.7] & 0.053\\
\midrule
Subfamily: Ectatomminae &  10.2 & 44.5 & 0.23 & [$-$81.1, 101.6] & 0.820\\
Subfamily: Formicinae   &   7.5 & 38.4 & 0.20 & [$-$71.3,  86.3] & 0.846\\
Subfamily: Myrmeciinae  &  50.6 & 51.0 & 0.99 & [$-$54.0, 155.2] & 0.329\\
Subfamily: Myrmicinae   &  51.4 & 46.1 & 1.11 & [$-$43.3, 146.1] & 0.275\\
\midrule
Body size (centred)     &  13.3 & 18.1 & 0.73 & [$-$23.8,  50.4] & 0.470\\
Colony size (centred)   &  11.3 & 25.8 & 0.44 & [$-$41.5,  64.2] & 0.663\\
Body size $\times$ Colony size & 9.4 & 32.8 & 0.29 & [$-$57.9,  76.8] & 0.776\\
\bottomrule
\end{tabular}
\begin{tablenotes}
\footnotesize
\item $N = 39$ colonies; one row per colony.  OLS fitted with
  \texttt{statsmodels} using heteroscedasticity-consistent standard errors.
\item \textbf{Variable definitions:} `Body size (centred)' =
  log\textsubscript{10} worker mass in mg, centred at the assemblage mean
  ($\bar{x} = 0$ on the centred scale); `Colony size (centred)' =
  log\textsubscript{10} total colony worker count, centred similarly.  The
  response is total days of colony observation (the survival window), an
  imperfect colony-level proxy for individual worker lifespan.
\item[\emph{Note}] The body-size\,$\times$\,colony-size interaction
  $p = 0.78$ (coefficient test, this table) and $p = 0.60$ (F-test vs.\ the
  reduced model) are both consistent with the null that colony size does not
  moderate the size--longevity benefit.
\end{tablenotes}
\end{threeparttable}
\end{table}

The one signal of an environmental role is directional rather than
conclusive.  The protective effect of body size is stronger in the field
(HR\,=\,0.67) than in the laboratory (HR\,=\,0.73, 95\,\%\,CI [0.45--1.21],
$p = 0.22$) --- a 1.28-fold difference on the log-hazard scale
($|\hat\beta|$: 0.40 vs.\ 0.31) --- and field and laboratory survival
indices are positively but not significantly correlated across species
(Spearman $\rho = 0.41$, $p = 0.088$, $N = 18$).
Taken together --- a robust main effect, two null moderators (colony size,
temperature), a weak but significant foraging-rate interaction, and a null
thermal interaction --- the evidence remains consistent with intrinsic
physiology as the most parsimonious account of the size--longevity rule
(inferred by exclusion of extrinsic-buffering and colony-dilution
alternatives), while acknowledging a modest foraging-context qualification;
it cannot formally rule out unmeasured confounders at $N = 18$ species.

%% ---- Table 3: Cox model comparison --------------------------
\begin{table}[htbp]
\caption{Cox proportional-hazards model comparison (Claim~1, H2b).
  All models include cluster-robust standard errors
  (cluster\,=\,\texttt{cohort\_id}), subfamily fixed effects (5\,levels,
  Dolichoderinae reference), and daily maximum temperature.
  $^{a}$LRT $\chi^2(1)$ vs.\ M0; $^{b}$LRT $\chi^2(1)$ vs.\ M0$_{\rm fr}$.
  \label{tab:cox_compare}}
\begin{threeparttable}
\begin{tabular}{lrrrrrr}
\toprule
\textbf{Model} & \textbf{k} & \textbf{log-lik} & \textbf{AIC} &
\textbf{Concordance} & $\boldsymbol{\chi^2(1)}$ & \textbf{\textit{p}}\\
\midrule
M0: \texttt{logw\_c + maxt + C(SF)}                  & 6 & $-12264.88$ & 24541.8 & 0.637 & ---          & ---   \\
M0$_{\rm fr}$: M0 $+$ \texttt{fr\_c}                 & 7 & $-12264.67$ & 24543.3 & 0.637 & $0.42^{a}$   & 0.52  \\
M2: M0$_{\rm fr}$ $+$ \texttt{logw\_c}$\times$\texttt{fr\_c} & 8 & $-12261.63$ & 24539.3 & 0.636 & $6.08^{b}$ & 0.014 \\
\bottomrule
\end{tabular}
\begin{tablenotes}
\footnotesize
\item M0 is the primary model for Claims~1 and~3 (H1, H3, H7; see Methods).
  M2 (here denoted as in Methods \S2.2) adds the body-size\,$\times$\,foraging-rate
  interaction for H2b.  Adding \texttt{fr\_c} alone
  (M0\,$\to$\,M0$_{\rm fr}$) yields no AIC improvement
  ($\Delta$AIC\,=\,$+$1.6, LRT $p = 0.52$); the interaction term in M2
  is significant vs.\ M0$_{\rm fr}$ ($\Delta$AIC\,=\,$-$4.1, $p = 0.014$)
  and vs.\ M0 ($\Delta$AIC\,=\,$-$2.5).
\end{tablenotes}
\end{threeparttable}
\end{table}

\subsection{Claim 2 --- Mortality trajectory is associated with circadian niche,
  not body size}
\label{sec:claim2}

\begin{figure}[htbp]
\centering
\includegraphics[width=0.85\textwidth]{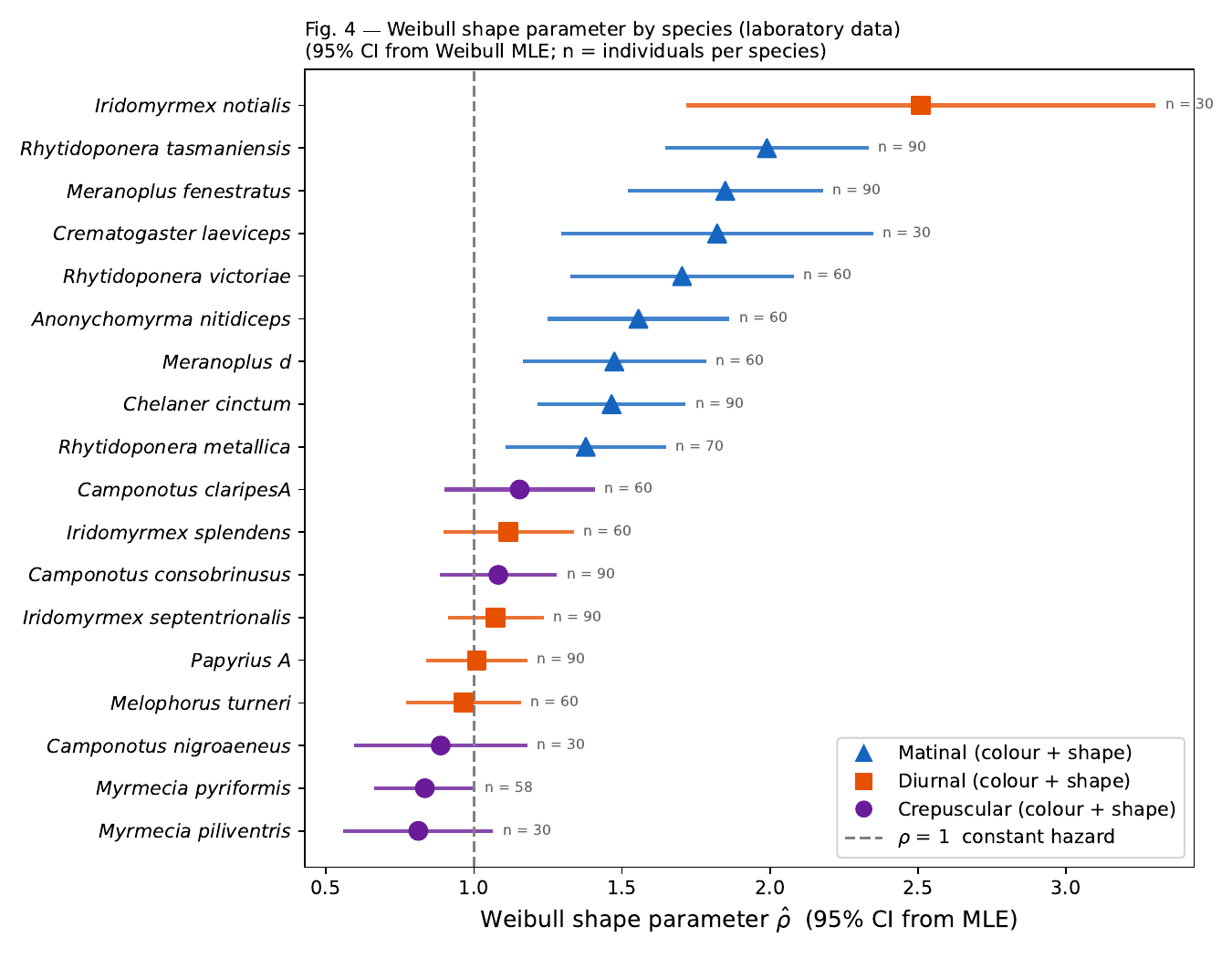}
\caption{\textbf{Weibull shape parameters $\rho$ by species, estimated from
  laboratory survival data (Claim~2, H4, H5).}  Species are sorted in
  ascending order of $\hat\rho_{\rm lab}$.  Horizontal bars are 95\,\%
  confidence intervals from maximum-likelihood Weibull fits to
  individual-level laboratory survival; sample sizes ($n$ individuals per
  species) are annotated to the right of each estimate.  The dashed vertical
  line at $\rho = 1$ separates frailty selection ($\rho < 1$) from senescence
  ($\rho > 1$).  Points are coloured \emph{and} shaped by circadian activity niche:
  matinal (blue triangle~$\blacktriangle$), diurnal (orange square~$\blacksquare$),
  crepuscular (purple circle~$\bullet$) --- dual encoding makes the grouping
  legible in greyscale and for colour-vision deficiencies.  Fourteen of 18 species have
  $\rho > 1$ (actuarial senescence).  The
  distribution of $\rho$ across niches illustrates the circadian structuring
  of trajectory: matinal species concentrate at higher $\rho$ values
  (Kruskal--Wallis $H = 9.35$, $p = 0.009$; matinal vs.\ crepuscular
  $p = 0.002$).
  \label{fig:weibull}}
\end{figure}

Senescence is the rule rather than the exception: 14 of 18 species display
senescent laboratory mortality trajectories ($\rho > 1$; mean
$\rho = 1.37$, median $\rho = 1.27$; Table~\ref{tab:weibull},
Fig.~\ref{fig:weibull}).  The explanation is not body size: across species,
$\rho$ is uncorrelated with body mass (Spearman $\rho = -0.25$, $p = 0.32$),
a key dissociation --- the trait that predicts duration (Claim~1) does not
predict the shape of ageing.

Trajectory is instead associated with circadian activity regime.  $\rho$ differs
significantly among circadian niches (Kruskal--Wallis $H = 9.35$,
$p = 0.009$), with matinal species senescing most steeply and crepuscular
species most shallowly; the matinal vs.\ crepuscular contrast is significant
($p = 0.002$), whereas matinal vs.\ diurnal is not ($p = 0.09$).  The latter
is blunted by a single diurnal outlier, \textit{Iridomyrmex notialis}, which
has the steepest trajectory in the dataset ($\rho = 2.51$).  The circadian
signal is therefore best described as a graded matinal-to-crepuscular
continuum rather than a clean three-way separation.

A formal variance partitioning of $\rho_{\rm lab}$ (three OLS models:
circadian niche only, subfamily only, and combined; $N = 18$ species)
quantified the collinearity directly.  Adjusted $R^2$ for circadian
niche alone was $0.33$ and for subfamily alone $0.31$, yet the full
model achieved only $R^2_{\rm adj} = 0.19$ ($F_{6,11} = 1.67$,
$p = 0.22$), with a shared overlap fraction of $-0.45$.  The strongly
negative overlap term indicates that circadian niche and subfamily are
largely redundant predictors at $N = 18$ species, consistent with
Cram\'{e}r's $V = 0.85$ (Supplementary Table~\ref{tab:vpRho}).

Finally, the field accelerates senescence uniformly: mortality trajectories
are steeper in the field than in the laboratory in 17 of 18 species (Wilcoxon
signed-rank $p < 0.001$; Table~\ref{tab:delta_rho}).
In a sensitivity analysis restricted to the 15 species with independently
estimated field $\rho$ (excluding three species whose field estimate is based
on a pooled-assemblage fallback), 14 of 15 show $\Delta\rho > 0$ (Wilcoxon
$p < 0.001$), indicating that the result is not driven by the fallback
procedure.
This amplification is near-universal rather than size-dependent: it is
consistent with the \emph{shape} of the trajectory being an
intrinsic, species-typical property whose steepness is nonetheless amplified
under field conditions.  \textit{Rhytidoponera} (Ectatomminae) carry
among the highest laboratory trajectories ($\rho_{\rm lab} = 1.38$--1.99;
Fig.~\ref{fig:weibull}), i.e.\ steep senescence is already present before any
thermal effect is considered.

%% Fig 5 — Weibull shape by species

%% ---- Table 6: Weibull shape parameter per species -----------
\begin{table}[htbp]
\caption{Weibull shape parameter $\rho_{\rm lab}$ estimated from laboratory
  mortality data, with species ordered by descending $\rho_{\rm lab}$
  (Claims~2, H4, H5).  Circadian niche is assigned from peak foraging hour
  (M\,=\,matinal, D\,=\,diurnal, C\,=\,crepuscular).  The shape parameter
  classifies the trajectory: $\rho > 1$, senescence (rising hazard);
  $\rho \approx 1$, constant hazard; $\rho < 1$, frailty selection (declining
  hazard).  $\lambda_{\rm lab}$ is the Weibull scale parameter (characteristic
  lifetime, in days); \textbf{Median (d)} is the model-implied median lifespan.
  95\,\% CIs are from the maximum-likelihood Weibull fit
  (\texttt{lifelines} \texttt{WeibullFitter}); the parenthetical value in the
  \textbf{Pattern} column is $n$, the number of individuals assayed per species.
  The two bottom rows summarise the assemblage: mean\,/\,median $\rho_{\rm lab}$,
  and the count of senescent species ($\rho_{\rm lab} > 1$) out of 18.
  \label{tab:weibull}}
\begin{threeparttable}
\small
\begin{tabular}{llrrcrrl}
\toprule
\textbf{Species} & \textbf{Subfamily} & \textbf{Niche} &
$\boldsymbol{\rho_{\rm lab}}$ & \textbf{95\,\% CI} &
$\boldsymbol{\lambda_{\rm lab}}$ &
\textbf{Median (d)} & \textbf{Pattern}\\
 & & & & & & & ($n$)\\
\midrule
\textit{Iridomyrmex notialis}        & Dolichoderinae & D & 2.51 & [1.72, 3.30] & 22.5 & 19.4 & Senescence (30)\\
\textit{Rhytidoponera tasmaniensis}  & Ectatomminae   & M & 1.99 & [1.65, 2.33] & 58.5 & 48.7 & Senescence (90)\\
\textit{Meranoplus fenestratus}      & Myrmicinae     & M & 1.85 & [1.52, 2.17] & 83.6 & 68.6 & Senescence (90)\\
\textit{Crematogaster laeviceps}     & Myrmicinae     & M & 1.82 & [1.30, 2.34] & 35.5 & 29.0 & Senescence (30)\\
\textit{Rhytidoponera victoriae}     & Ectatomminae   & M & 1.70 & [1.33, 2.07] & 55.6 & 44.8 & Senescence (60)\\
\textit{Anonychomyrma nitidiceps}    & Dolichoderinae & M & 1.56 & [1.25, 1.86] & 8.6  &  6.8 & Senescence (60)\\
\textit{Meranoplus diversus}         & Myrmicinae     & M & 1.47 & [1.17, 1.78] & 89.3 & 69.6 & Senescence (60)\\
\textit{Chelaner cinctum}            & Myrmicinae     & M & 1.46 & [1.22, 1.71] & 69.4 & 54.0 & Senescence (90)\\
\textit{Rhytidoponera metallica}     & Ectatomminae   & M & 1.38 & [1.11, 1.64] & 68.2 & 52.3 & Senescence (70)\\
\textit{Camponotus claripes}~A       & Formicinae     & C & 1.15 & [0.91, 1.40] & 44.3 & 32.2 & Senescence (60)\\
\textit{Iridomyrmex splendens}       & Dolichoderinae & D & 1.12 & [0.90, 1.33] & 19.8 & 14.3 & Senescence (60)\\
\textit{Camponotus consobrinus}      & Formicinae     & C & 1.08 & [0.89, 1.27] & 98.7 & 70.3 & $\approx$Constant (90)\\
\textit{Iridomyrmex septentrionalis} & Dolichoderinae & D & 1.07 & [0.92, 1.23] & 21.8 & 15.5 & $\approx$Constant (90)\\
\textit{Papyrius} A                  & Dolichoderinae & D & 1.01 & [0.85, 1.17] & 22.6 & 15.7 & $\approx$Constant (90)\\
\textit{Melophorus turneri}          & Formicinae     & D & 0.96 & [0.78, 1.15] & 28.4 & 19.5 & $\approx$Constant (60)\\
\textit{Camponotus nigroaeneus}      & Formicinae     & C & 0.89 & [0.60, 1.17] & 82.5 & 54.6 & Frailty (30)\\
\textit{Myrmecia pyriformis}         & Myrmeciinae    & C & 0.83 & [0.67, 1.00] & 18.4 & 11.8 & Frailty (58)\\
\textit{Myrmecia piliventris}        & Myrmeciinae    & C & 0.81 & [0.57, 1.06] & 34.3 & 21.8 & Frailty (30)\\
\midrule
\multicolumn{3}{l}{\textbf{Mean / median}} & 1.37\,/\,1.27 & & & &\\
\multicolumn{3}{l}{\textbf{Senescent ($\rho > 1$)}} & 14/18 & & & &\\
\bottomrule
\end{tabular}
\begin{tablenotes}
\footnotesize
\item Circadian niche: M\,=\,matinal, D\,=\,diurnal, C\,=\,crepuscular.
  95\,\% CI: confidence interval from the Weibull MLE summary;
  $n$ (in Pattern column): number of individuals assayed.
  Kruskal--Wallis across niche groups: $H = 9.35$, $p = 0.009$.
  Matinal vs.\ crepuscular: $p = 0.002$; matinal vs.\ diurnal: $p = 0.09$.
  Spearman correlation of $\rho$ vs.\ body mass: $\rho = -0.25$, $p = 0.32$.
\end{tablenotes}
\end{threeparttable}
\end{table}

%% ---- Table 7: Field amplification of senescence (H6) --------
\begin{table}[htbp]
\caption{Field amplification of mortality trajectory (H6): $\Delta\rho =
  \rho_{\rm field} - \rho_{\rm lab}$ per species.  Positive $\Delta\rho$
  indicates steeper senescence in the field.  Species are grouped by
  circadian niche.  $\dagger$\,=\,field $\rho$ from pooled AFT fallback.
  \label{tab:delta_rho}}
\begin{threeparttable}
\small
\begin{tabular}{llrrrr}
\toprule
\textbf{Species} & \textbf{Niche} & $\boldsymbol{\rho_{\rm lab}}$ &
$\boldsymbol{\rho_{\rm field}}$ & $\boldsymbol{\Delta\rho}$ &
\textbf{Direction}\\
\midrule
\textit{Meranoplus fenestratus}      & Matinal & 1.85 & 3.28 & $+1.43$ & $\uparrow$\\
\textit{Rhytidoponera metallica}     & Matinal & 1.38 & 0.54 & $-0.84$ & $\downarrow$\\
\textit{Rhytidoponera tasmaniensis}  & Matinal & 1.99 & 2.94 & $+0.95$ & $\uparrow$\\
\textit{Rhytidoponera victoriae}     & Matinal & 1.70 & 3.59 & $+1.88$ & $\uparrow$\\
\textit{Anonychomyrma nitidiceps}    & Matinal & 1.56 & 2.12 & $+0.57$ & $\uparrow$\\
\textit{Chelaner cinctum}            & Matinal & 1.46 & 1.34 & $+0.13$ & $\uparrow$\\
\textit{Crematogaster laeviceps}\tnote{$\dagger$}  & Matinal & 1.82 & 5.65 & $+3.83$ & $\uparrow$\\
\textit{Meranoplus diversus}         & Matinal & 1.47 & 2.60 & $+1.13$ & $\uparrow$\\
\midrule
\textit{Iridomyrmex notialis}\tnote{$\dagger$}    & Diurnal & 2.51 & 5.65 & $+3.14$ & $\uparrow$\\
\textit{Iridomyrmex septentrionalis} & Diurnal & 1.07 & 4.69 & $+3.62$ & $\uparrow$\\
\textit{Iridomyrmex splendens}       & Diurnal & 1.12 & 3.99 & $+2.87$ & $\uparrow$\\
\textit{Melophorus turneri}          & Diurnal & 0.96 & 5.15 & $+4.19$ & $\uparrow$\\
\textit{Papyrius} A                  & Diurnal & 1.01 & 3.79 & $+2.78$ & $\uparrow$\\
\midrule
\textit{Camponotus claripes}~A       & Crepuscular & 1.15 & 4.89 & $+3.74$ & $\uparrow$\\
\textit{Camponotus consobrinus}      & Crepuscular & 1.08 & 2.85 & $+1.77$ & $\uparrow$\\
\textit{Camponotus nigroaeneus}\tnote{$\dagger$}  & Crepuscular & 0.89 & 5.65 & $+4.76$ & $\uparrow$\\
\textit{Myrmecia piliventris}        & Crepuscular & 0.81 & 4.47 & $+3.66$ & $\uparrow$\\
\textit{Myrmecia pyriformis}         & Crepuscular & 0.83 & 1.25 & $+0.42$ & $\uparrow$\\
\midrule
\multicolumn{4}{l}{\textbf{Species with $\Delta\rho > 0$}} & 17/18 &\\
\multicolumn{5}{l}{\textbf{Wilcoxon signed-rank test vs.\ 0}} & $p < 0.001$\\
\bottomrule
\end{tabular}
\begin{tablenotes}
\footnotesize
\item[$\dagger$] Field $\rho$ estimated from pooled assemblage AFT fallback.
\end{tablenotes}
\end{threeparttable}
\end{table}

\subsection{Claim 3 --- Thermal risk is nonlinear and taxonomically
  patterned; body size offers no protection}
\label{sec:claim3}

Temperature raises daily hazard by 3.6\,\% per \textdegree C in the
main-effects model (HR\,=\,1.036, 95\,\%\,CI [1.016--1.057], $p < 0.001$;
Table~\ref{tab:cox_main}) --- equivalent to a ${\sim}19\,\%$ increase across
5\,\textdegree C of warming.  But this linear summary conceals a strongly
nonlinear reality.  A piecewise spline with a breakpoint at
$T^* = 20$\,\textdegree C minimises AIC across the full search grid
($T^* \in \{15$--$34\}$\,\textdegree C; $\Delta$AIC\,=\,$-38$ vs.\ the
linear model; Davies permutation $p < 0.001$, $B = 1{,}000$;
Table~\ref{tab:spline}, Fig.~\ref{fig:spline}A).  Below 20\,\textdegree C the
marginal thermal hazard is steep (HR\,=\,1.216 per \textdegree C, 95\,\%\,CI
[1.117, 1.325], $p < 0.001$); above 20\,\textdegree C it collapses to a
near-plateau (HR\,=\,1.016 per \textdegree C, 95\,\%\,CI [1.004, 1.027],
$p = 0.006$).  Because 65.7\,\% of cohort-day observations fall above this
threshold (mean \texttt{maxt}\,=\,22.8\,\textdegree C), most of the realised
thermal variation lies within the attenuated regime.  The pattern is
consistent with behavioural buffering --- foragers retreating to the nest
once surface temperatures exceed the activity optimum, which would cap
realised thermal exposure \citep{Andrew2013, Youngsteadt2023} --- but
this remains a hypothesis inferred from the mortality pattern rather than a
direct behavioural measurement (we develop the interpretation in
\S\ref{sec:claim3}).
External evidence for nest retreat under heat is well documented in genera
present in this assemblage.
\textit{Iridomyrmex purpureus} (closely related to our three
\textit{Iridomyrmex} species) maintains body temperatures below trail
temperatures during peak heat, with air temperatures at weather-station
height at least 10\,\textdegree C below simultaneous ground-surface
temperatures \citep{Andrew2013}.  Under this calibration, our breakpoint
$T^* = 20$\,\textdegree C (air) corresponds to a ground-surface
temperature of ${\approx}30$--35\,\textdegree C --- the range at which
\textit{Rhytidoponera convexa} (our genus \textit{Rhytidoponera}) reaches
its published upper foraging limit \citep{Jayatilaka2011}, and at which
\textit{Myrmecia} species transition from unimodal to bimodal foraging to
avoid peak heat \citep{Jayatilaka2011}.  These correspondences are consistent
with the hypothesis that the plateau marks the onset of widespread
behavioural retreat rather than a hard physiological ceiling
\citep{Parr2022, Andrew2013, Jayatilaka2011}, the lineages with ready access
to nest retreats being those most buffered against gradual warming; we did
not, however, observe retreat behaviour directly, and confirming the
mechanism would require behavioural assays under progressive warming.

%% ---- Table 8: Piecewise spline model (H8) -------------------
\begin{table}[htbp]
\caption{Piecewise spline model of the temperature--hazard function
  (H8, Claim~3), in two panels.  \textbf{Upper panel:} model comparison;
  the best breakpoint $T^* = 20$\,\textdegree C was selected by minimising AIC
  across $T^* \in \{15, 16, \ldots, 34\}$\,\textdegree C.
  \textbf{Lower panel:} marginal thermal hazard within each spline segment.
  $\Delta$AIC, change in Akaike information criterion relative to the linear
  baseline (negative values favour the piecewise model);
  HR/\textdegree C, hazard ratio per 1\,\textdegree C increase in daily
  maximum temperature; $T^*$, estimated breakpoint;
  LRT, likelihood-ratio test; a dash (---) denotes a quantity that is not
  applicable (the linear model is the reference).
  \label{tab:spline}}
\begin{threeparttable}
\begin{tabular}{lrrrl}
\toprule
\textbf{Model} & \textbf{AIC} & $\boldsymbol{\Delta}$\textbf{AIC} &
\textbf{LRT \textit{p}} & \textbf{Note}\\
\midrule
Linear (M0, reference) & 24{,}542 & 0 & --- & Baseline\\
Piecewise spline, $T^* = 20$\,\textdegree C &
  24{,}504 & $-$38 & $< 0.001$ & Best breakpoint\\
\bottomrule
\end{tabular}

\medskip

\begin{tabular}{lrrrl}
\toprule
\textbf{Spline segment} & \textbf{HR/\textdegree C} & \textbf{95\,\% CI} &
\textbf{\textit{p}} & \textbf{Interpretation}\\
\midrule
Below $T^* = 20$\,\textdegree C  & 1.216 & [1.117, 1.325] & $< 0.001$ & Steep thermal hazard\\
Above $T^* = 20$\,\textdegree C  & 1.016 & [1.004, 1.027] & 0.006 & Near-plateau\\
\bottomrule
\end{tabular}
\begin{tablenotes}
\footnotesize
\item $\Delta$AIC\,=\,$-37.7$ vs.\ linear model (reported as $-38$ in text);
  Davies permutation test (permute \texttt{maxt} within cohorts;
  $B = 1{,}000$): $p < 0.001$ (0 of 1{,}000 permuted datasets reached
  $\Delta$AIC\,$\leq -37.7$).  The below-$T^*$ HR is the \texttt{maxt}
  coefficient; the above-$T^*$ HR and its 95\,\% CI are the linear
  combination ($\beta_{\texttt{maxt}} + \beta_{(\texttt{maxt}-T^*)_+}$) with
  cluster-robust variance propagation.
  65.7\,\% of cohort-day observations fall above $T^* = 20$\,\textdegree C
  (mean \texttt{maxt}\,=\,22.8\,\textdegree C).
\end{tablenotes}
\end{threeparttable}
\end{table}

This buffer is not shared equally.  For between-species comparisons of
thermal sensitivity, the phylogenetically appropriate primary framework is
PGLS.  Pagel's $\lambda = 0$ (LRT $p = 1.00$) indicates no detectable
phylogenetic signal in species-level thermal sensitivities, so PGLS reduces
to OLS; nonetheless, using species as the unit of analysis is more
conservative than pooling all worker-days.  In this species-level analysis,
the \textit{Rhytidoponera} lineage (Ectatomminae) shows directionally
elevated thermal sensitivity: PGLS
$\hat\beta_\text{Ectatomm} = +0.048$/\textdegree C ($p = 0.081$,
$N = 14$ species; Table~\ref{tab:pgls}).  The result does not reach
$\alpha = 0.05$ --- reflecting limited statistical power at $N = 14$
species --- but the direction is consistent across all specifications.

Corroborating this genus-level signal, the individual-level Cox model
draws on the full worker-day dataset ($N = 2{,}363$ observations;
1{,}826 deaths) and thus has substantially greater power: the joint
subfamily\,$\times$\,temperature interaction is significant
(LRT $\chi^2(4) = 12.0$, $p = 0.017$; Table~\ref{tab:subfamily_temp},
Fig.~\ref{fig:spline}), and only the Ectatomminae contrast --- here
represented entirely by \textit{Rhytidoponera} --- shows a significant
positive term: elevated thermal sensitivity of 5\,\% per \textdegree C
(HR\,=\,1.05, 95\,\%\,CI [1.01--1.09], $p = 0.015$) above the assemblage
response.  During extreme heat events (maxt\,$> 30$\,\textdegree C),
\textit{Rhytidoponera} experience a 66\,\% excess hazard (HR\,=\,1.66,
95\,\%\,CI [1.11--2.48], $p = 0.014$).  Together, the PGLS direction and
individual-level significance point to elevated thermal sensitivity
observed in the \textit{Rhytidoponera} lineage (Ectatomminae) --- not a
phylogenetically conserved clade-wide pattern, but a consistent
genus-level signal visible across both analytical frameworks.
Body size provides no protection at any temperature (Claim~1, H3).

%% ---- Table 9: Subfamily × temperature interaction (H9) ------
\begin{table}[htbp]
\caption{Subfamily\,$\times$\,temperature interaction model coefficients (H9,
  Claim~3).  Reference: Dolichoderinae\,$\times$\,temperature.  Panel~A:
  continuous \texttt{maxt\_c} interaction; Panel~B: binary heat-event
  indicator (\texttt{heat30}: maxt\,$> 30$\,\textdegree C).  Both models
  include body size and subfamily fixed effects.
  \label{tab:subfamily_temp}}
\begin{threeparttable}
\textbf{Panel A: Continuous temperature interaction}\\[2pt]
\begin{tabular}{lrrrrl}
\toprule
\textbf{Subfamily interaction} & $\boldsymbol{\beta}$ & \textbf{HR} &
\textbf{95\,\% CI} & & \textbf{\textit{p}}\\
\midrule
maxt\_c (Dolichoderinae reference slope) &
  $+$0.014 & 1.014 & [0.986, 1.044] & & 0.328\\
maxt\_c $\times$ Ectatomminae &
  $+$0.048 & 1.050 & [1.009, 1.091] & $\star$ & 0.015\\
maxt\_c $\times$ Formicinae &
  $+$0.023 & 1.023 & [0.974, 1.075] & & 0.367\\
maxt\_c $\times$ Myrmeciinae &
  $+$0.013 & 1.013 & [0.966, 1.064] & & 0.589\\
maxt\_c $\times$ Myrmicinae &
  $+$0.019 & 1.019 & [0.962, 1.078] & & 0.524\\
\bottomrule
\end{tabular}

\medskip

\textbf{Panel B: Extreme heat events (maxt $> 30$\,\textdegree C)}\\[2pt]
\begin{tabular}{lrrrl}
\toprule
\textbf{Coefficient} & \textbf{HR} & \textbf{95\,\% CI} & &
\textbf{\textit{p}}\\
\midrule
heat30 (Dolichoderinae reference) & 1.207 & [0.918, 1.587] & & 0.177\\
heat30 $\times$ Ectatomminae & 1.656 & [1.106, 2.479] & $\star$ & 0.014\\
heat30 $\times$ Formicinae & 1.131 & [0.700, 1.828] & & 0.616\\
heat30 $\times$ Myrmeciinae & 1.130 & [0.651, 1.963] & & 0.663\\
heat30 $\times$ Myrmicinae & 0.899 & [0.433, 1.863] & & 0.774\\
\bottomrule
\end{tabular}
\begin{tablenotes}
\footnotesize
\item $\star$ Significant at $\alpha = 0.05$.  Reference subfamily:
  Dolichoderinae.  HR, hazard ratio; CI, cluster-robust 95\,\% confidence
  interval; \texttt{maxt\_c}, daily maximum temperature mean-centred at the
  assemblage average (22.8\,\textdegree C); \texttt{heat30}, binary indicator
  equal to 1 when daily maximum temperature exceeds 30\,\textdegree C and 0
  otherwise.  In Panel A the \texttt{maxt\_c} row is the reference-group
  (Dolichoderinae) temperature slope from the interaction model; the
  assemblage-average temperature effect (HR\,=\,1.036) is reported for the
  pooled model M0 in Table~\ref{tab:cox_main}.
  \emph{Joint likelihood-ratio test} for the subfamily\,$\times$\,temperature
  interaction (Panel A): $\chi^2(4) = 12.0$, $p = 0.017$ (vs.\ the
  no-interaction model).  Only the \textit{Rhytidoponera}/Ectatomminae
  contrast departs significantly in either panel; all other subfamily
  interactions are non-significant.
\end{tablenotes}
\end{threeparttable}
\end{table}

A parallel PGLS for the body-size log-hazard
signal ($\hat\beta_\text{logw}$ per species, $N = 14$) yielded
$\hat\lambda = 0.06$ (LRT $p = 0.90$), indicating that the
size--longevity pattern likewise lacks detectable phylogenetic structure
(Supplementary Table~\ref{tab:pgls_size}).

%% ============================================================
\section{Discussion}
\label{sec:discussion}

Our central result is a \textbf{three-axis reorganisation}: no single
predictor accounts for all three components of worker mortality risk, and
body size --- the canonical predictor --- accounts for only one of them.
This reorganises a size-centric framework into a three-axis one, with direct
consequences for how we predict ant responses to environmental change.  We
emphasise that the \emph{conceptual contribution} --- evidence consistent
with duration, senescence trajectory, and thermal vulnerability being
decoupled and associated with different predictors --- is the primary
advance.  The specific identity of the most vulnerable lineage
(\textit{Rhytidoponera}, Ectatomminae; Claim~3) is a secondary, provisional
finding that motivates a testable hypothesis rather than a settled conclusion.

\subsection{Duration is associated with intrinsic physiology (Claim 1)}

The convergence of a robust main effect with two null moderators (colony
size, temperature) and a weak but significant foraging-rate interaction
(LRT $p = 0.014$) narrows the mechanistic window for the
size--longevity rule.  The pattern is more consistent with intrinsic physiology --- mass-specific
metabolic scaling and reactive-oxygen-species production
\citep{Brown2004, Kramer2021} --- than with body size acting primarily as an
extrinsic buffer \citep{Lighton1994, Bujan2016}, though foraging intensity
provides a modest contextual qualification.  Crucially, the CI on the
size\,$\times$\,temperature interaction [0.98, 1.03] is inconsistent with any
thermal moderation of the size benefit larger than
$\approx$\,3\,\%/\textdegree C.  With 2,363 cohort-day observations this test
has ${\approx}97\,\%$ power to detect an interaction as small as
HR\,=\,1.05 (\S\ref{sec:discussion}, Limitations): the null is therefore
positive evidence against ecologically meaningful thermal buffering by body
size, not a failure of detection.  We cannot formally exclude a weaker
buffering effect, or one operating at finer temporal or spatial resolution
than captured by \texttt{maxt}.  The directional field amplification (1.28\texttimes\ on the log-hazard scale) is most
parsimoniously read as extrinsic mortality filtering through an intrinsic
size advantage, rather than body size buffering any specific extrinsic stressor.
Frailty models confirm that this is a between-species, not within-species,
pattern (Table~\ref{tab:frailty}): intraspecific body-size variation in
this assemblage is too narrow to detect an independent individual-level
effect, and the HR\,=\,0.67 should be interpreted as a species-level
comparative signal alongside other between-species analyses.

\subsection{Trajectory is associated with circadian regime, not size (Claim 2)}

The dissociation between body size and trajectory (Spearman $\rho = -0.25$,
$p = 0.32$) is the conceptual hinge of the paper: duration and trajectory are
different axes with different predictors.  Trajectory shape is associated with circadian
niche, with matinal species senescing fastest.  Matinal foragers, active as
surface temperatures rise steeply in the early morning, may experience a more
variable and physiologically taxing thermal environment, accelerating
senescence.  This parallels `pace-of-life' syndromes described in vertebrates
\citep{Ricklefs2002, Debecker2018}, but expressed here at the level of
species-specific activity regimes.  We are cautious about over-reading the
three-way niche contrast: the matinal--diurnal difference is not individually
significant and one diurnal species departs strongly from its group, so we
describe the signal as a graded matinal-to-crepuscular continuum.

\subsection{Thermal risk is nonlinear --- and the buffer is taxon-specific
  (Claim 3)}

The thermal plateau above $T^* = 20$\,\textdegree C is the most novel result
of this study.  It departs from the monotonically rising limb of the
thermal performance curve expected below the activity optimum
\citep{Martin2008, Deutsch2008, Arnoldi2025}: rather than mortality
continuing to climb with temperature, hazard saturates.  The pattern is
consistent with behavioural buffering (thermoregulation) --- workers
retreating below ground when surface temperatures exceed foraging optima
would impose a ceiling on realised thermal exposure
\citep{Andrew2013, Jayatilaka2011}.  This hypothesis is supported by the
correspondence between $T^* = 20$\,\textdegree C (air) and published
foraging-limit temperatures for this assemblage (see Results), but is not
directly demonstrated here.  The plateau onset
corresponds to a daily maximum air temperature of 20\,\textdegree C, at
which point ground surface temperatures already substantially exceed air
temperature --- consistent with conditions that trigger nest retreat in
Australian ants, which restrict field activity well below their physiological
thermal maxima \citep{Jayatilaka2011} --- a pattern documented more broadly
across ant assemblages globally \citep{Nascimento2022}.

Importantly, the plateau does not imply invulnerability above
20\,\textdegree C.  First, the hazard ratio remains above 1 throughout ---
thermal mortality continues, merely attenuated.  Second, elevated thermal
sensitivity in the \textit{Rhytidoponera} lineage (excess hazard 5\,\% per
\textdegree C) is consistent with some lineages being unable to fully execute
behavioural retreat, possibly because their matinal schedule exposes them to
morning temperature ramps before retreat is possible --- a hypothesis
requiring direct behavioural testing.  Third, extreme heat events
($> 30$\,\textdegree C) produce large excess mortality (66\,\% for
\textit{Rhytidoponera}) that likely overwhelms behavioural buffering
altogether.

\subsection{Synthesis: circadian regime and lineage identity --- collinear
  climate-relevant axes}

The three claims converge on a single reframing (Fig.~\ref{fig:synthesis}).
Body size, the trait most
often used as a proxy for risk, predicts only \emph{duration}; it is silent
on \emph{trajectory} and on \emph{thermal vulnerability}.  Those two
climate-relevant axes are associated with circadian activity regime and
lineage identity, respectively --- axes that are strongly collinear in this
assemblage (Cram\'{e}r's $V = 0.85$) and intersect in one lineage.
Importantly, circadian regime and lineage identity cannot be fully
disentangled in this assemblage.  The three
signals are not equally robust.  The body-size effect (HR\,=\,0.67,
$p = 0.002$) and the null size\,$\times$\,temperature interaction (CI
[0.98, 1.03]) are the most firmly established results; the Kruskal--Wallis
circadian-niche effect on $\rho$ ($p = 0.009$) and the
\textit{Rhytidoponera}/Ectatomminae Cox interaction ($p = 0.015$) are well
supported at the individual level but rest on $N = 18$ species or a single
genus; and the PGLS signal for this lineage ($p = 0.081$) is suggestive
only.  With that ranking in mind, three signals converge on
\textit{Rhytidoponera} (Ectatomminae): steep senescent
trajectories (Claim~2), a directionally elevated but non-significant PGLS
thermal coefficient (Claim~3, $\hat\beta = +0.048$/\textdegree C,
$p = 0.081$, $N = 14$ species; power limited to detect genus-level signals at
this sample size), and a matinal activity regime --- three signals that
\emph{converge} without individually constituting strong evidence.  The
individual-level Cox model
corroborates this with greater statistical power ($p = 0.015$, $N =
2{,}363$ cohort-days), but the PGLS --- the appropriate framework for
between-species comparisons --- remains the primary test; together they
indicate elevated thermal sensitivity observed in the \textit{Rhytidoponera}
lineage (Ectatomminae) rather than a phylogenetically conserved clade-wide
pattern.  The genus \textit{Rhytidoponera} is widespread across eastern and
south-eastern
Australia and ecologically important as predators and seed dispersers
\citep{Lubertazzi2010, Parr2007}.  Because the mean daily maximum already
sits at the plateau onset, the most consequential climate pathway is not
gradual warming of the mean but the rising frequency of extreme-heat days
($> 30$\,\textdegree C), where behavioural buffering fails and where
\textit{Rhytidoponera} are most exposed \citep{IPCC2021, Herold2018,
PerkinsKirkpatrick2020}.

\subsection{Implications: what changes if risk is three-dimensional}

Treating mortality risk as a single, size-indexed axis is incomplete.  Our
results show that purely size-based vulnerability frameworks are
insufficient: they capture only the \emph{duration} axis and can be improved
by incorporating the two additional axes identified here (senescence
trajectory and thermal vulnerability structured by circadian regime and
lineage identity).  Trait-based vulnerability frameworks routinely assess
thermal tolerance \citep{Deutsch2008, Diamond2017}; our results show that
such indices are, by construction, silent on the two axes that structure
climate exposure here (size\,$\times$\,temperature $p = 0.72$).  A framework
that substituted circadian activity regime and lineage identity for body size
would have flagged matinal \textit{Rhytidoponera} (Ectatomminae) as most
exposed; a size-based index would not, and would in fact rank them as
intermediate.

Two consequences follow for forecasting.  First, because the
temperature--hazard function plateaus near the assemblage mean
($T^* = 20$\,\textdegree C; mean \texttt{maxt}\,=\,22.8\,\textdegree C; a
result firmly supported by $\Delta$AIC\,=\,$-38$ and Davies $p < 0.001$),
gradual warming of the \emph{mean} should have limited effect on realised
mortality, whereas the rising \emph{frequency} of extreme-heat days
($> 30$\,\textdegree C) --- the regime in which behavioural buffering fails
and \textit{Rhytidoponera} carry a 66\,\% excess hazard (Cox HR\,=\,1.66
[1.11, 2.48], individual-level only; to be treated as preliminary pending
broader sampling) --- is the operative climate
variable \citep{PerkinsKirkpatrick2020, IPCC2021}.  Vulnerability projections
built on mean-temperature anomalies will therefore understate risk for
heat-sensitive lineages.  Second, because the most exposed lineage
(\textit{Rhytidoponera}) is a dominant predator and seed disperser in
south-eastern Australia \citep{Lubertazzi2010, Parr2007}, lineage-specific
heat mortality has a plausible pathway to ecosystem function ---
myrmecochory and predation --- that an assemblage-average analysis would
obscure.  We frame both as actionable hypotheses rather than firm
projections: at $N = 18$ species, with circadian niche and lineage identity
strongly collinear (Cram\'{e}r's $V = 0.85$), the priority is to redirect
monitoring and trait-collection effort, not to issue lineage-specific risk
verdicts.

\subsection{Limitations}
\label{sec:limitations}

\begin{enumerate}
  \item \textbf{Sample size and genus-driven signals.}  Eighteen species
    limit power for subfamily-level inference.  The thermal signal attributed
    to Ectatomminae in the models is driven entirely by three
    \textit{Rhytidoponera} species, and the corresponding PGLS coefficient
    ($p = 0.081$) does not cross the conventional significance threshold at
    this $N$.  Claims~2 and~3 should therefore be read as \emph{provisional
    patterns} rather than settled conclusions: replication across other
    matinal, ground-foraging lineages in different biogeographic regions is
    needed before general statements about subfamily-level thermal
    vulnerability can be made.
  \item \textbf{Temperature resolution and microhabitat mismatch.}
    \texttt{maxt} is a daily maximum recorded at a weather station
    ${\sim}1$\,km from the study site, which can diverge from microhabitat
    temperatures by 10--20\,\textdegree C depending on nest depth and
    substrate \citep{Andrew2013}.  Surface-foraging species (many
    Ectatomminae) are likely more exposed to station-level \texttt{maxt} than
    subterranean or nocturnal foragers, so our thermal-hazard estimates may
    be partly confounded with foraging guild, and the breakpoint
    $T^* = 20$\,\textdegree C may reflect guild-level heterogeneity as well
    as a physiological threshold.
  \item \textbf{Field vs.\ laboratory $\rho$.}  Laboratory conditions
    (controlled temperature, ad libitum food) differ markedly from the field,
    and several field $\rho$ estimates rest on few cohorts; $\Delta\rho$
    therefore reflects both genuine extrinsic acceleration and environmental
    confounding, and should be read as a qualitative direction, not a precise
    magnitude.  For three species with fewer than three cohorts
    (\textit{Camponotus nigroaeneus}, \textit{Crematogaster laeviceps},
    \textit{Iridomyrmex notialis}), $\rho_{\rm field}$ was estimated from a
    pooled fallback model (flagged $\dagger$); a sensitivity analysis
    excluding these three species confirmed that 14 of 15 remaining species
    show $\Delta\rho > 0$ (Wilcoxon $p < 0.001$), indicating that the
    conclusion of field amplification of senescence is not driven by the
    fallback estimates.
  \item \textbf{Collinearity (foraging rate).}  The foraging-rate\,$\times$\,
    body-size interaction carries moderate collinearity
    (VIF\,$\approx$\,4), so the corresponding interaction
    (LRT $p = 0.014$) should be read cautiously.
  \item \textbf{Collinearity (circadian niche and subfamily).}  Circadian
    niche and subfamily identity are strongly collinear in this assemblage:
    all three \textit{Rhytidoponera} (Ectatomminae) are matinal and both
    \textit{Myrmecia} (Myrmeciinae) are crepuscular (Cram\'{e}r's $V = 0.85$,
    $\chi^2(8) = 25.9$, $p = 0.001$).  Claims 2 and 3 cannot be fully
    separated with $N = 18$ species, so the attribution to \emph{circadian
    niche} (Claim~2) versus \emph{subfamily identity} (Claim~3) is
    operationally indistinguishable at this scale.  Two of our conclusions are
    nonetheless robust to this collinearity: the body-size HR (an
    interspecific effect, independent of niche classification) and the null
    size\,$\times$\,temperature interaction.  Two are tentative: the specific
    attribution of trajectory structuring to circadian niche rather than
    lineage identity, and the attribution of thermal vulnerability to
    Ectatomminae as a subfamily rather than to matinal \textit{Rhytidoponera}
    specifically.  Disentangling these will require cross-phylogenetic
    sampling --- multiple matinal lineages drawn from different subfamilies.
    Reviewers may ask: \emph{``How can you separate circadian niche from
    lineage identity?''}  The honest answer is: we cannot, with $N = 18$ from
    a single assemblage.  We present the attribution to circadian niche
    (Claim~2) and to lineage identity (Claim~3) as two complementary framings
    of the same collinear signal, and propose the separation as the key
    empirical question for future cross-phylogenetic work.
  \item \textbf{Hypothesis-generating scope.}  This study is
    \emph{confirmatory} for Claim~1 (body-size duration effect: HR\,=\,0.67,
    $p = 0.002$, robust to multiple sensitivity analyses) and for the null
    size\,$\times$\,temperature interaction (CI [0.98, 1.03]).  It is
    \emph{hypothesis-generating} for Claims~2 and~3: the circadian-niche
    structuring of trajectory and the thermal excess in \textit{Rhytidoponera}
    are consistent, directional signals that motivate targeted replication, but
    cannot be treated as confirmed at the subfamily or clade level from a
    single 18-species assemblage.  Reviewers should assess Claims~2 and~3
    by the standard of pattern discovery, not by the standard of mechanistic
    proof.
  \item \textbf{Statistical power at $N = 18$ (simulation-based).}
    We estimated retrospective power via Monte Carlo simulation
    (3,000 replicates per design point; \texttt{block0\_power/power\_analysis.py};
    fixed seed\,=\,42).
    \begin{enumerate}[label=(\roman*)]
      \item \textit{Spearman $\rho_{\rm lab}$ vs.\ body mass ($N = 18$,
        two-tailed $\alpha = 0.05$).}  Estimated power at the observed
        $|\rho| = 0.25$ is $15\,\%$; the minimum detectable correlation at
        $80\,\%$ power is $|\rho| = 0.65$.  The null result ($\rho = -0.25$,
        $p = 0.32$) therefore cannot exclude true associations as large as
        $|\rho| \approx 0.60$.
      \item \textit{PGLS Ectatomminae thermal coefficient ($N = 14$:
        three Ectatomminae vs.\ eleven others; $\hat\lambda \approx 0$,
        so a $t$-test is a valid proxy).}  Power at Cohen's $d = 1.0$ is
        $41\,\%$ (one-sided); the minimum detectable effect at $80\,\%$
        power is $d \approx 1.8$.  The observed $p = 0.081$ is fully
        consistent with a genuine but underpowered elevation.
      \item \textit{Cox size\,$\times$\,temperature interaction
        ($N = 2,\!363$ cohort-days; analytical Wald test).}
        The interaction SE\,=\,0.013 affords $80\,\%$ power to detect
        HR\,$\geq 1.04$; HR\,=\,1.05 is detectable at $97\,\%$ power.
        The null result (HR\,=\,1.00 [0.98,\,1.03]) therefore effectively
        rules out biologically meaningful size\,$\times$\,temperature
        interactions.
      \item \textit{Cox body-size main effect ($N = 2,\!363$).}
        Power at the observed HR\,=\,0.67 is $87\,\%$, confirming the
        robustness of Claim~1.
    \end{enumerate}
    We include tests with $<50\,\%$ power for completeness and
    directionality; they should not be read as decisive evidence.
  \item \textbf{Phylogenetic power.}  The PGLS robustness check used
    $N = 14$ of 18 species (four failed to converge due to near-zero
    intraspecific body-size variance); Pagel's $\lambda = 0$ precludes
    phylogenetically informed weighting.  The parallel PGLS for the
    body-size log-hazard signal also returned $\hat\lambda = 0.06$
    ($p = 0.90$), confirming that neither size--longevity nor
    size--temperature sensitivity shows detectable phylogenetic structure.
    Elevated thermal sensitivity in \textit{Rhytidoponera} ($p = 0.015$ in
    the individual-level Cox) should therefore be treated as a genus-level
    hypothesis rather than a confirmed clade-wide pattern.
\end{enumerate}

%% ============================================================
\section{Conclusion}
\label{sec:conclusion}

Worker mortality risk in ants is organised along three axes, each associated
with a different predictor.  Body size predicts longevity \emph{duration},
consistent with intrinsic physiology, independently of colony size and
temperature (with a weak foraging-rate qualification).  The \emph{trajectory}
of mortality --- the rate of senescence --- is associated with circadian
activity regime, not body size.  \emph{Thermal risk} is nonlinear,
plateauing above 20\,\textdegree C in a pattern consistent with behavioural
buffering, with elevated thermal sensitivity observed in the
\textit{Rhytidoponera} lineage (Ectatomminae; Pagel's $\lambda = 0$) that
persists above the plateau and intensifies during extreme heat.  By showing
that the trait predicting how long a worker lives says nothing about how it
ages or how it dies from heat, these results reframe the size--longevity
paradigm.  The practical consequence is concrete: predicting ant responses to
warming requires tracking the frequency of extreme-heat events rather than
shifts in mean temperature, and weighting circadian regime and lineage
identity --- not body size --- when ranking which taxa, and which ecosystem
functions, are most at risk.

%% ============================================================
\section*{Acknowledgements}

[To be completed at submission; must include land acknowledgement and any
technical assistance.]

\section*{Conflict of Interest}

The authors declare no competing interests.

\section*{Data and Code Availability}

All raw data are available via Dryad \citep[doi: 10.5061/dryad.j0zpc86s4]{Riskas2026}.
Analysis code is available at \url{https://github.com/[FILL-BEFORE-SUBMISSION]}
as a fully reproducible Python pipeline (\texttt{source\_files/}), with a
frozen archive deposited on Zenodo
(\url{https://doi.org/[FILL-BEFORE-SUBMISSION]}).  The repository includes a
\texttt{README} documenting execution order and expected outputs.

%% ============================================================
\bibliography{paperA}

%% ============================================================
%% SUPPLEMENTARY TABLES
%% ============================================================

\clearpage
\section*{Supplementary Tables}
\setcounter{table}{0}
\renewcommand{\thetable}{S\arabic{table}}

%% ---- Supplementary Table S1: Schoenfeld residuals -------------------
\begin{table}[h!]
\caption{Proportional hazards assumption test via Schoenfeld
  residuals for the main Cox model M0
  (\texttt{logw\_c + maxt + C(SF)}, $N = 2{,}363$ cohort-day observations,
  1{,}826 events, cluster\,=\,\texttt{cohort\_id}).  Test statistic and
  two-sided $p$-value from
  \texttt{lifelines.statistics.proportional\_hazard\_test} using the
  rank-transformed time axis (Grambsch--Therneau test, rank variant); the
  rank transform reduces sensitivity to the baseline hazard shape and
  improves power against smooth time-varying alternatives.
  \texttt{logw\_c}: log\textsubscript{10} body mass, mean-centred;
  \texttt{maxt}: daily maximum temperature (\textdegree C).  No covariate
  shows evidence of a time-varying hazard ratio at $\alpha = 0.05$,
  supporting the PH assumption for M0.
  \label{tab:schoenfeld}}
\begin{threeparttable}
\small
\begin{tabular}{lrr}
\toprule
\textbf{Covariate} & \textbf{Test statistic} & \textbf{\textit{p}}\\
\midrule
Body size (\texttt{logw\_c})   & 1.854 & 0.173\\
Max.\ temperature (\texttt{maxt}) & 1.410 & 0.235\\
Subfamily: Ectatomminae & 0.292 & 0.589\\
Subfamily: Formicinae   & 0.323 & 0.570\\
Subfamily: Myrmeciinae  & 0.569 & 0.451\\
Subfamily: Myrmicinae   & 1.529 & 0.216\\
\bottomrule
\end{tabular}
\begin{tablenotes}
\footnotesize
\item All $p \geq 0.17$; the PH assumption is not violated for any
  primary predictor.
\end{tablenotes}
\end{threeparttable}
\end{table}

%% ---- Supplementary Table S2: Species-frailty robustness check ------
\begin{table}[h!]
\caption{Species-frailty robustness check for H1.
  Three nested Cox models are compared to assess whether the body-size hazard
  ratio (HR\,=\,0.67 in M0) is robust when species identity is controlled.
  M0\_spp adds species fixed effects (18 dummies absorbing all
  between-species variation); M0\_strat uses species as strata, allowing
  species-specific baseline hazards.  Both variants yield a directionally
  consistent but statistically non-significant HR, indicating that the M0
  coefficient reflects a predominantly interspecific size--longevity
  relationship.  \label{tab:frailty}}
\begin{threeparttable}
\small
\begin{tabular}{llrr}
\toprule
\textbf{Model} & \textbf{Species control} & \textbf{HR$_\text{logw}$} & \textbf{\textit{p}}\\
\midrule
M0             & Subfamily fixed effects (ref) & 0.672 & 0.002\\
M0\_spp        & Species fixed effects (18 dummies) & 0.770 & 0.471\\
M0\_strat      & Stratified Cox by species & 0.710 & 0.354\\
\bottomrule
\end{tabular}
\begin{tablenotes}
\footnotesize
\item HR direction is consistent across all models ($< 1$, indicating
  larger workers live longer).  Non-significance in M0\_spp and M0\_strat
  reflects narrow intraspecific variation in \texttt{logw\_mass} rather than
  absence of a size effect.  All models cluster standard errors by cohort.
\end{tablenotes}
\end{threeparttable}
\end{table}

%% ---- Supplementary Table S3: PGLS thermal-sensitivity test ---------
\begin{table}[h!]
\caption{Phylogenetic generalised least-squares (PGLS)
  analysis of subfamily differences in thermal sensitivity (primary
  analysis for between-species comparisons, Claim~3).  Response variable:
  per-species $\hat{\beta}_\text{maxt}$ from individual-species Cox models
  ($N = 14$ of 18 species converged).  Predictor: subfamily (reference =
  Dolichoderinae).  Phylogenetic covariance matrix derived from the
  ultrametric \texttt{mr\_phylo.tre} tree \citep{Riskas2026}, scaled by
  Pagel's $\hat{\lambda} = 0$ (LRT $p = 1.00$; no detectable phylogenetic
  signal), so PGLS reduces to OLS.  \label{tab:pgls}}
\begin{threeparttable}
\small
\begin{tabular}{lrrrr}
\toprule
\textbf{Term} & \textbf{Coef.} & \textbf{SE} & \textbf{\textit{t}} & \textbf{\textit{p}}\\
\midrule
Intercept (Dolichoderinae)  & 0.018 & 0.016 & 1.16 & 0.277\\
Ectatomminae                & 0.048 & 0.024 & 1.97 & 0.081\\
Formicinae                  & 0.024 & 0.024 & 0.98 & 0.355\\
Myrmeciinae                 & 0.034 & 0.036 & 0.95 & 0.367\\
Myrmicinae                  & 0.035 & 0.024 & 1.44 & 0.185\\
\midrule
\multicolumn{4}{l}{Joint $F(4,9) = 1.09$} & 0.416\\
\bottomrule
\end{tabular}
\begin{tablenotes}
\footnotesize
\item Pagel's $\hat{\lambda} = 0.00$ (LRT $\chi^2(1) = 0.00$, $p = 1.00$).
  Four species excluded due to convergence failure in species-specific Cox
  models (\textit{C.\ nigroaeneus}, \textit{C.\ laeviceps},
  \textit{I.\ notialis}, \textit{M.\ piliventris}).  The
  \textit{Rhytidoponera}/Ectatomminae PGLS coefficient is directionally
  consistent with the individual-level Cox result (HR\,=\,1.05,
  $p = 0.015$) but does not reach $\alpha = 0.05$ at $N = 14$ species.
\end{tablenotes}
\end{threeparttable}
\end{table}

%% ---- Supplementary Table S4: PGLS body-size longevity signal -------
\begin{table}[h!]
\caption{Phylogenetic generalised least-squares (PGLS) for the body-size
  log-hazard signal (A1 robustness check).  Response variable: per-species
  $\hat{\beta}_\text{logw}$ from individual-species Cox models
  (\texttt{logw\_c + maxt}; $N = 14$ converged).  Model: intercept-only
  PGLS to obtain the assemblage-level phylogenetically corrected mean.
  Pagel's $\hat{\lambda} = 0.06$ (LRT $\chi^2(1) = 0.02$, $p = 0.90$):
  no detectable phylogenetic signal, supporting retention of the M0 Cox
  analysis as the primary model.
  \label{tab:pgls_size}}
\begin{threeparttable}
\small
\begin{tabular}{lrrl}
\toprule
\textbf{Term} & $\hat{\lambda}$ & \textbf{LRT \textit{p}} & \textbf{Conclusion}\\
\midrule
Intercept (assemblage $\beta_\text{logw}$) & 0.06 & 0.90 &
  No phylogenetic structure; M0 retained\\
\bottomrule
\end{tabular}
\begin{tablenotes}
\footnotesize
\item Four species excluded due to convergence failure (\textit{C.\ nigroaeneus},
  \textit{C.\ laeviceps}, \textit{I.\ notialis}, \textit{M.\ piliventris}).
  $\hat\lambda < 0.3$ indicates that the body-size mortality signal does not
  co-vary appreciably with phylogenetic distance, supporting the
  non-phylogenetic Cox approach as the primary analysis.
\end{tablenotes}
\end{threeparttable}
\end{table}

%% ---- Supplementary Table S5: Variance partitioning rho_lab --------
\begin{table}[h!]
\caption{Variance partitioning of $\rho_{\rm lab}$ between circadian niche
  and subfamily (A3 robustness check; $N = 18$ species).  Adjusted $R^2$
  for three OLS models: niche-only, subfamily-only, and combined.
  Negative shared fraction confirms high collinearity between predictors
  (Cram\'{e}r's $V = 0.85$), consistent with the non-significant full
  model ($F_{6,11} = 1.67$, $p = 0.22$).  \label{tab:vpRho}}
\begin{threeparttable}
\small
\begin{tabular}{lrr}
\toprule
\textbf{Component} & $R^2_{\rm adj}$ & \textbf{Interpretation}\\
\midrule
Circadian niche only     & 0.33 & Unique niche contribution\\
Subfamily only           & 0.31 & Unique subfamily contribution\\
Full model (niche + SF)  & 0.19 & Combined (not additive due to collinearity)\\
Shared (overlap)         & $-$0.45 & Redundancy; collinearity confirmed\\
\bottomrule
\end{tabular}
\begin{tablenotes}
\footnotesize
\item Variance partitioning follows \citet{Legendre2012}.  The strongly
  negative shared fraction ($-0.45$) arises when predictors are highly
  collinear: adding a redundant predictor to an already-saturated model
  reduces adjusted $R^2$.  This is not a modelling artefact but a
  quantitative expression of the Cram\'{e}r's $V = 0.85$ collinearity
  reported in the Limitations.
\end{tablenotes}
\end{threeparttable}
\end{table}

\end{document}